\documentclass[twocolumn,epjc3]{svjour3}          

\RequirePackage[T1]{fontenc}

\smartqed  

\RequirePackage{mathptmx}      
\RequirePackage{flushend}
\RequirePackage[numbers,sort&compress]{natbib}
\RequirePackage[colorlinks,citecolor=blue,urlcolor=blue,linkcolor=blue]{hyperref}

\usepackage{amsmath}
\usepackage{graphicx,subfigure}
\usepackage{amssymb}
\usepackage{bm}

\usepackage{ulem}
\usepackage{color}

\journalname{Eur. Phys. J. C}

\begin{document}


\title{Nucleon strangeness polarization from $\Lambda/\bar\Lambda$ hyperon production in polarized proton-proton collision at RHIC}


\author{Xiaonan Liu\thanksref{addr1}
        \and
        Bo-Qiang Ma\thanksref{e1,addr1,addr2,addr3} 
}

\thankstext{e1}{e-mail: mabq@pku.edu.cn}

\institute{School of Physics and State Key Laboratory of Nuclear Physics and Technology, Peking University, Beijing 100871, China\label{addr1}
          \and
          Collaborative Innovation Center of Quantum Matter, Beijing, China\label{addr2}
          \and
          Center for High Energy Physics, Peking University, Beijing 100871, China\label{addr3}
}

\date{Received: date / Accepted: date}

\maketitle

\begin{abstract}
We calculate the inclusive production of a polarized $\Lambda$ or $\bar{\Lambda}$ hyperon
from the single longitudinally polarized proton and proton ($pp$) collision at RHIC.
By comparing the data reported by the STAR Collaboration, we find that this process
is sensitive to the polarization of strange and antistrange quarks of the proton in
the experimental range. By introducing asymmetric coefficients with the minimization
of $\chi^2$, we further identify the asymmetry of the polarized strange-antistrange
quarks in the proton sea and find that the first moment is $\Delta s \approx -0.025\pm 0.019$ for strange quark
and $\Delta\bar s \approx -0.001\pm 0.012$ for antistrange quark, with central values agreeing with the light-cone meson-baryon fluctuation model prediction, the recent lattice QCD determination of strangeness polarization and results from a global QCD analysis given by the Jefferson Lab Angular Momentum (JAM) Collaboration.
\end{abstract}

\section{Introduction}
\label{intro}

The spin content of the proton has received extensive attentions since the so-called
 ``proton spin crisis''~\cite{Ashman:1987hv,Ashman:1989ig}.
There have been significant progress on spin distributions of light flavor valence quarks
both experimentally and theoretically. Recently, some efforts have been made on the
polarization of the light flavor sea quarks~\cite{Tian:2017xul,Tian:2017qwk} and the
quark-antiquark asymmetry of helicity distributions in nucleon sea~\cite{Liu:2018mio}.
Studies of the momentum distribution asymmetry between strange and
antistrange~\cite{Brodsky:1996hc,Signal:1987gz,Burkardt:1991di,Holtmann:1996be,Sufian:2018cpj}
are of significance for the study of nonperturbative quantum chromodynamics~(QCD) with fruitful
progress~\cite{Ma:1997gh,Barone:1999yv,Ding:2004ht,Ding:2004dv,Wakamatsu:2004pd,Alwall:2005xd,
Ding:2005ub,Gao:2005gj,Hao:2005dw,Bourrely:2007if,Diehl:2007uc,Gao:2007ht,Gao:2008ch,Zhou:2009mx,
Chi:2014xba,Hobbs:2014lea,Wakamatsu:2014asa,Vega:2015hti,Du:2017nzy}.
However, the understanding of helicity distributions for strange-antistrange quark
pairs $\Delta s(x)$ and $\Delta \bar{s}(x)$ is far from satisfactory and needs further studies.
Researches show that studies of spin-transfer reactions
~\cite{Owens:1986mp,Stratmann:1992gu,Kotzinian:1997vd,deFlorian:1998ba,
Ma:2000uu,Boros:2000ex,Ma:2001na,Ellis:2002zv,Xu:2002hz,Xu:2004es,Xu:2005ru,Chen:2007tm,Zhou:2010vm}
and measurements of the polarized structure function~\cite{Ashman:1987hv,Ashman:1989ig,
Adams:1994zd, Abe:1994cp,Abe:1997cx,Adams:1997tq,Adeva:1998vv, Airapetian:2006vy},
can provide new insights into the nucleon spin structure and yield a better understanding
of the hadronization process.
The $\Lambda$ baryon plays a special role in this respect since it has a rather simple
spin structure in the naive quark parton model~\cite{GellMann:1964nj}, as well as that
its self-analyzing parity violating decays~\cite{Lee:1957qs} make polarization measurements
experimentally feasible.

The STAR experiment at Relativistic Heavy Ion Collider~(RHlC)
~\cite{Harrison:2003sb,Harrison:2002es,Courant:2003ad} is carrying
out a spin physics program in high-energy polarized proton-proton
collisions at $\sqrt{s}=200\,\mathrm{GeV}$ and $\sqrt{s}=500\,\mathrm{GeV}$
in order to gain a deeper insight into the nucleon spin structure and dynamics
of the proton. Theoretical investigations
~\cite{deFlorian:1998ba,Boros:2000ex,Ma:2001na,Xu:2002hz,Xu:2004es,Xu:2005ru,Chen:2007tm,Zhou:2010vm}
suggest that the spin transfer process of the $\Lambda/\bar\Lambda$ production in the
singly polarized proton-proton collision provides a platform to extract polarized parton
distribution functions of the proton and polarized quark fragmentation functions of quarks
into $\Lambda/\bar{\Lambda}$ hyperon. The experiments measuring the polarization of $\Lambda$
hyperon produced in singly polarized $pp$ collisions~\cite{Abelev:2009xg,Adam:2018kzl} may
provide information about not only the inclusive production of hadrons but also the
strange and antistrange quark polarizations inside the proton.

Compared to lepton-induced reaction experiments, the corresponding cross section
of the hadron-hadron collision is larger and the luminosity of the incoming proton
beams can be made higher. Moreover, there are many subprocesses that may lead to
the final state $\Lambda$ in singly polarized $pp$ collisions due to the rich
internal structure of the proton, such as $qq\rightarrow qq$,
$qg\rightarrow qg$, $gg\rightarrow gg$, etc., which may produce large partonic spin asymmetries.
Therefore productions of $\Lambda$ and $\bar{\Lambda}$ are sensitive to polarized
parton distributions in the proton, making STAR ideal to explore quark and gluon polarization.

In this work, we perform the calculation of longitudinal spin transfers to $\Lambda$ and $\bar{\Lambda}$
hyperons produced in polarized proton-proton collisions at RHIC.
By comparing recent data~\cite{Adam:2018kzl} reported by the STAR Collaboration, we find
that this process is sensitive to the polarization of strange and antistrange quarks
of the proton in the experimental range. Such result agrees with the conclusions in previous
studies~\cite{Xu:2005ru,Chen:2007tm,Zhou:2010vm}.
By introducing asymmetric coefficients with
 the minimization of $\chi^2$ in fitting data, we further identify the asymmetry of
 the polarized strange-antistrange quarks in the proton and find that the first moment is $\Delta s \approx -0.025\pm 0.019$ for strange quark and
$\Delta \bar s \approx -0.001\pm 0.012$ for antistrange quark, with central values agreeing with the light-cone meson-baryon fluctuation model prediction~\cite{Brodsky:1996hc}, the recent lattice QCD determination of strangeness polarization~\cite{QCDSF:2011aa} and results from a global QCD analysis given by the Jefferson Lab Angular Momentum (JAM) Collaboration~\cite{Ethier:2017zbq}.

The reminder of this paper
is organized as follows. Sect.~\ref{cross_section} provides general
formulae for production cross section. As the $\Lambda/\bar\Lambda$
 hyperon production process in singly polarized $pp$ collisions contains
many different hard subprocesses, Sect.~\ref{analysis} elucidates kinematic
regions and identifies the major subprocess. We then calculate fragmentation
functions for quarks to $\Lambda$ in Sect.~\ref{FFs}. Numerical results of the
spin transfer to $\Lambda/\bar\Lambda$ are shown in Sect.~\ref{results} with
considerations of the asymmetry of polarized strange-antistrange quarks in
the proton, together with discussions and conclusions.
 A brief summary is given in Sect.~\ref{Summary}.

\section{Theoretical calculation}
\subsection{The general formulae for cross section}
\label{cross_section}

We consider the single inclusive production process $\overrightarrow{p}p \rightarrow \overrightarrow{\Lambda} X$, where a polarized $\Lambda/\bar\Lambda$ is produced from the single longitudinally polarized $pp$ collision at RHIC. The longitudinal spin transfer to $\Lambda/\bar{\Lambda}$ hyperon produced in polarized proton-proton collisions is defined as~\cite{Owens:1986mp}
\begin{equation*}
A^{\Lambda/\bar\Lambda} = E\frac{d^{3}\Delta\sigma}{d^{3}p} / E\frac{d^{3}\sigma}{d^{3}p},
\label{eq:spin_trans}
\end{equation*}
where $\mathrm{d}\Delta\sigma$ and $\mathrm{d}\sigma$ stand for the polarized and unpolarized cross sections, respectively.
The definition and derivation of the parton-parton scattering contribution to the single-particle inclusive cross section can be found in Refs.~\cite{Berman:1971xz,Ellis:1973nb}. By using factorization theorem~\cite{Collins:1987pm}, the differential cross-section for high-transverse momentum distributions has been computed to next-to-leading order accuracy in perturbative QCD~\cite{Aversa:1988vb,deFlorian:2002az,Jager:2002xm}.
In this way, such cross sections can be expressed as convolutions of perturbatively calculable partonic spin-transfer cross sections~\cite{Owens:1986mp,Stratmann:1992gu,Boros:2000ex} with certain sets of parton distributions and fragmentation functions at the scale $Q$, just as follows,
\begin{equation}
\begin{split}
  E\frac{d^{3}\Delta\sigma}{d^{3}p} (\mathrm{AB} \rightarrow \mathrm{C}+\mathrm{X})
  =
&  \sum_{abcd}
  \int_{\bar x_{a}}^1 dx_a
   \int_{\bar x_{b}}^1 dx_b   \Delta f^{\mathrm{A}}_{a}(x_a,Q^2)  \\
   &  f^{\mathrm{B}}_{b}(x_b, Q^2 ) \Delta D^{\mathrm{C}}_{c}(z_c, Q^2) \frac{1}{\pi z_c}  \\
  &  \frac{d\Delta\hat\sigma}{d\hat{t}}
 (ab\rightarrow cd),
 \label{eq:cross_sec}
\end{split}
\end{equation}
with
\begin{equation}
\begin{split}
    z_c  &= \frac{x_{\mathrm{T}}}{2x_b} e^{-\eta} + \frac{x_{\mathrm{T}}}{2x_a} e^\eta, \\
     \bar x_{b} &= \frac{x_{\mathrm{a}}x_{\mathrm{T}} e^{-\eta}}{2x_a-x_{\mathrm{T}} e^\eta}, \\
    \bar x_{a} &=  \frac{x_{\mathrm{T}} e^\eta}{2-x_{\mathrm{T}} e^{-\eta}},
    \label{eq:momen_frac}
\end{split}
\end{equation}
here $\eta$ and $p_{\mathrm{T}}$ are rapidity and transverse momentum, and $x_{\mathrm{T}} = 2p_{\mathrm{T}} /\sqrt{s}$ is the transverse momentum counterpart. We denote the momenta of incoming protons and the produced hyperon by $P_\mathrm{A}$, $P_\mathrm{B}$ and $P_\mathrm{C}$, respectively. $x_a$, $x_b$ and $z_c$ are corresponding  momentum fractions carried by partons $a$, $b$ and $c$, and $\bar x_{a}$ and $\bar x_{b}$ are  corresponding lower limits of integration.

In Eq.~(\ref{eq:cross_sec}), $\Delta f^{\mathrm{A}}_{a}(x_a,Q^2)$ represents the polarized distribution function of parton $a$ in proton $\mathrm{A}$ and $f^{\mathrm{B}}_{b}(x_b, Q^2)$ the unpolarized one of parton $b$ in proton $\mathrm{B}$, at the scale $Q$.
$\Delta D^{\mathrm{C}}_{c}(z_c, Q^2)$ is the polarized fragmentation function of
parton $c$ into baryon $\mathrm{C}$, and in our case $\mathrm{C}=\Lambda$. $\Delta\mathrm{d}\hat\sigma/\mathrm{d}\hat{t}$, the difference of cross sections at the parton level between the two processes $a^\uparrow +b\rightarrow c^\uparrow + d$ and
$a^\downarrow +b\rightarrow c^\uparrow + d$, can be found in, for example, Ref.~\cite{Owens:1986mp}.
The summation in Eq.~(\ref{eq:cross_sec}) runs over all possible
parton-parton subprocesses, $qq \rightarrow qq$,
$qg\rightarrow qg$, $q\bar{q}\rightarrow q\bar{q}$, etc..
 An analogous expression for unpolarized cross section is obtained by Eq.~(\ref{eq:cross_sec})
with the $\Delta$ symbol dropped throughout. In the calculation we adopt de~Florian-Sassot-Stratmann-Vogelsang~(DSSV)
polarized parton distribution functions~(PDFs)~\cite{deFlorian:2008mr,deFlorian:2009vb} as inputs to spin-dependent parton distributions of proton $A$ directly. With sufficient statistics, this set of PDFs not only provides the total up and down quarks distributions, $\Delta u+\Delta\bar{u}$ and
$\Delta d+\Delta\bar d$, and light sea quark polarizations, $\Delta\bar u$, $\Delta\bar d$, $\Delta\bar s$= $\Delta s$, but also gives an important constraint on the gluon polarization $\Delta g$. Considering that this set of PDFs adopts Martin-Stirling-Thorne-Watt~(MSTW) unpolarized PDFs~\cite{Martin:2009iq} as a reference set to avoid problems with the fundamental positivity constraint, we adopt the MSTW parametrization as unpolarized PDFs for proton $B$ in order to maintain self-consistency. The non-perturbative fragmentation functions for $\Lambda$ are given by the revised Gribov-Lipatov relation~\cite{Gribov:1971zn,Gribov:1972ri,Brodsky:1996cc,Barone:2000tx,Ma:2002ur} and the light-cone SU(6) quark-spectator-diquark model~\cite{Ma:1999gj,Ma:1999wp} adjusted by the Albino-Kniehl-Kramer~(AKK) parametrization~\cite{Albino:2005mv,Albino:2008fy} based on considerations following Ref.~\cite{Du:2017nzy}.

The Mandelstam variables at the parton level are given by
\begin{equation}
\begin{split}
  \hat{s} &=x_a x_b s,  \\
  \hat{t} &= - x_a p_{\mathrm{T}} \sqrt{s} e^{-\eta}/z_c, \\
  \hat{u} &= - x_b p_{\mathrm{T}} \sqrt{s} e^{\eta}/z_c,
  \label{eq:stu}
\end{split}
\end{equation}
here $s$ is the total center of mass energy.

It is obvious from Eq.~(\ref{eq:momen_frac}) that the lower
limits of momentum fraction of the initial state, $\bar{x}_{a}$
and $\bar{x}_{b}$, depend mainly on experimental conditions,
namely rapidity $\eta$ and transverse momentum $p_{\mathrm{T}}$.
The momentum fraction carried by the final state baryon, $z_{c}$,
is also related to the momentum fraction of the initial state. Note that
the magnitude of momentum fractions represents different regions of
momentum distributions, that is, the region where the momentum fraction
is smaller is mainly the sea region, and the region where the momentum
fraction is larger is the valence region. Such general relationships
can help us to analyze the dynamic range of $pp$ collisions in the
calculation and to obtain an understanding of the spin transfer in this process.

\subsection{Elucidate the kinematic regions and major subprocess}
\label{analysis}

The main aim of the spin transfer process to $\Lambda/\bar\Lambda$ hyperon produced
in $pp$ collisions is to extract the polarized quark and gluon distributions.
However, in the case of $pp$ collisions where many subprocesses can lead to final state
hyperons, we need to identify out the most contributing channel related to the polarized parton distributions.

We first check the relation between $\bar{x}_a$, the integral lower limit of $x_a$,
and $p_\mathrm{T}$ using Eq.~(\ref{eq:momen_frac}).
The small value of $p_\mathrm{T}$ is associated with $x_a$ in the sea part of the
polarized proton whereas typical values of $x_a$ corresponding to the valence region
correspond to a high $p_\mathrm{T}$, where the quarks are polarized much more
strongly, resulting in an asymmetry increase with $p_\mathrm{T}$.
Information about contribution from parton distributions of the polarized proton can also be tracked
from looking at the $x_a$ dependence of $\eta$.
Similarly, if there is a large positive $\eta$, this can be used to extract the
large $x_a$ part parton content of the proton.
The high $p_\mathrm{T}$ and large $\eta$ lead to dominating contribution to the
spin transfer to final state hyperons with the valence part of the polarized proton.
Hence it is reasonably to conclude that in large $p_{\mathrm{T}}$ region in $pp$
collisions, the fragmentation of quarks from the subprocesses $qg \rightarrow qg$
and $qq \rightarrow qq$ dominates. We expect that $u$ and $d$, particularly $u$, quark
fragmentation dominates the $\Lambda$ production in high $p_{\mathrm{T}}$ and large
rapidity region, as supported by previous studies~\cite{deFlorian:1998ba,Boros:2000ex,Ma:2001na,Xu:2002hz}.
The connection between $x_b$ and $x_a$ allows us to further determine the dynamic behavior.
It can be seen from Eq.~(\ref{eq:momen_frac}) that when $x_{a}$ is small, $x_{b}$ is relatively
large and mainly corresponds to the valence region, and as $x_{a}$ increases, the integral lower
limit of $x_{b}$ gradually expands to the sea region. Considering that the gluon is
distributing much greater over the sea region of the proton than quarks, the sum of valence
and sea quark contributions is suppressed at small $x_a$ and $x_b$. It means that the
subprocess $qg \rightarrow qg$ plays a major role than $qq \rightarrow qq$ case~\cite{Boros:2000ex,Ma:2001na}. It is
also argued in Ref.~\cite{Boros:2000ex} that the contribution from $gg \rightarrow gg$
falls off faster than that from $qg \rightarrow qg$ with increasing rapidity, since $g(x_{a})$
decreases faster than $q(x_{a})$ with increasing $x_{a}$. Therefore we can conclude that
the $qg \rightarrow qg$ subprocess plays the most important role in the $\Lambda/\bar\Lambda$
hyperon production of proton-proton collisions~\cite{deFlorian:1998ba,Boros:2000ex,Ma:2001na,Xu:2002hz},
so we only consider the spin transfer of this subprocess.


Based on the above considerations, the process of our calculation can be expressed as:
\begin{equation}
[\Delta q^{p}(x_{a})g^{p}(x_{b})+\Delta g^{p}(x_{a})q^{p}(x_{b})]\rightarrow [\Delta D_{q}^{\Lambda}(z_{c})+\Delta D_{g}^{\Lambda}(z_{c})].
\label{eq:qg}
\end{equation}
Specifically, this contains four subprocesses:
\begin{center}
\begin{enumerate}
\item[(1)] $\Delta q^{p_\mathrm{A}} g^{p_\mathrm{B}} \rightarrow \Delta q^{\Lambda} g$;
\item[(2)] $\Delta q^{p_\mathrm{A}} g^{p_\mathrm{B}} \rightarrow \Delta g^{\Lambda} q$;
\item[(3)] $\Delta g^{p_\mathrm{A}} q^{p_\mathrm{B}} \rightarrow \Delta q^{\Lambda} g$;
\item[(4)] $\Delta g^{p_\mathrm{A}} q^{p_\mathrm{B}} \rightarrow \Delta g^{\Lambda} q$.
\end{enumerate}
\end{center}
We write the detailed expressions in the following way:
\begin{equation}
\begin{split}
\Delta q^{p_\mathrm{A}}(x_{a}) g^{p_\mathrm{B}}(x_{b})\frac{d\sigma}{dt}(\hat{t},\hat{u}) [\Delta D_{q}^{\Lambda}(z_{c})+T^{\overrightarrow{q}g \rightarrow q\overrightarrow{g}}(y) \Delta D_{g}^{\Lambda}(z_{c})],    \\
\Delta g^{p_\mathrm{A}}(x_{a})q^{p_\mathrm{B}}(x_{b}) \frac{d\sigma}{dt}(\hat{u},\hat{t}) [T^{q\overrightarrow{g} \rightarrow \overrightarrow{q}g}(y) \Delta D_{q}^{\Lambda}(z_{c})+ \Delta D_{g}^{\Lambda}(z_{c})],      \\
\label{eq:qgall}
\end{split}
\end{equation}
where $q = u$, $d$, $s$, with $T^{\overrightarrow{q}g \rightarrow \overrightarrow{q}g}(y)=1$ omitted and $T^{q\overrightarrow{g} \rightarrow\overrightarrow{q}g}(y)=T^{\overrightarrow{q}g \rightarrow q\overrightarrow{g}}(y)=(1-(1-y)^2)/(1+(1-y)^2)$ being the polarization transfer factor for the two subprocesses, see Ref.~\cite{Xu:2002hz} for details. Due to that the fragmentation mechanism of the polarized gluon into $\Lambda$ is unclear, we take $\Delta D_{g}^{\Lambda} = 0$ in our calculation. Likewise, we use corresponding expressions analogy to Eq.(~\ref{eq:qgall}) for the longitudinal spin transfer to $\bar\Lambda$, by replacing $q$ with $\bar{q}$. There are several differences in the calculation of $\bar\Lambda$ compared to $\Lambda$. For $\bar\Lambda$ hyperon, the dominant subprocess should be $\Delta\bar{q}g \rightarrow \Delta\bar{q}g$, where we take $\Delta D_{\bar q}^{\bar\Lambda} =\Delta D_{q}^{\Lambda}$ according to CP symmetry. Due to the lack of
fragmentation functions for polarized sea quarks to $\Lambda$, we set $\Delta D_{q}^{\bar\Lambda}=\Delta D_{\bar q}^{\Lambda}=0$ in the calculation, based on same considerations as the gluon case. Different from the $\Lambda$ case, the polarized sea quarks play a main role in the process of spin transfer to $\bar{\Lambda}$, in which distributions of polarized sea quarks are relatively important.

\begin{figure}
\includegraphics[scale=0.5]{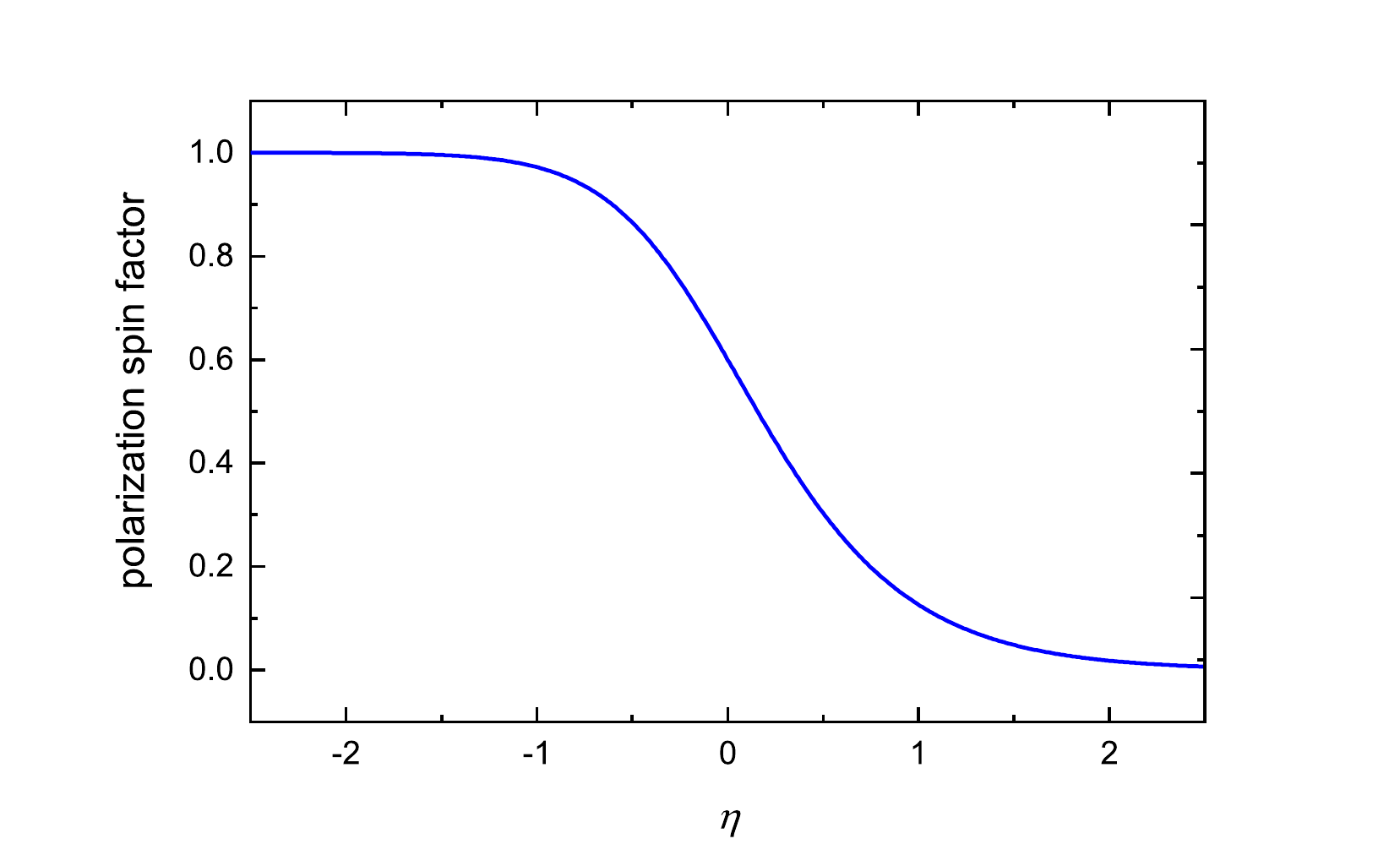}
\caption{The polarization transfer factor $T(\eta)$ versus rapidity $\eta$ of the
process $qg \rightarrow qg$. The solid curve represents the longitudinal polarization
transfer factor of the process $\Delta g \rightarrow \Delta q$.
The polarization transfer factor for $T^{\Delta {q} \rightarrow \Delta{q}}(\eta)=1$ is not presented in the picture.}
\label{fig:spin_T}
\end{figure}

For hadron-hadron collisions, each of the parton subprocess has its own polarization transfer factor.
For the polarization transfer factor $T(y)$, the
variable $y=p_\mathrm{B} \cdot (p_\mathrm{A}-p_\mathrm{C})/(p_\mathrm{A} \cdot p_\mathrm{B})$,
and in the center-of-mass frame $y=(1-\cos(\theta))/2$, where $\theta = 2\arctan(\exp(-\eta))$
is the angle between the incoming parton $a$ and the outgoing parton $c$.
Fig.~\ref{fig:spin_T} shows the polarization transfer factor $T(\eta)$ versus
rapidity $\eta$ of the process $qg \rightarrow qg$. It can be seen that for the process
 $q\overrightarrow{g} \rightarrow \overrightarrow{q}g$ the longitudinal polarization transfer factor
$T^{q\overrightarrow{g} \rightarrow \overrightarrow{q}g}(\eta)$ gradually tends to zero, as rapidity
increases. There is no dependence on rapidity for the longitudinal process
$\overrightarrow{q}g \rightarrow \overrightarrow{q}g$ with $T^{\overrightarrow{q}g \rightarrow \overrightarrow{q}g}(\eta)=1$.

\subsection{Fragmentation functions and AKK parametrization}
\label{FFs}

Fragmentation functions~(FFs) describe phenomena of
hadronization, i.e., how quarks or gluons transform into hadrons.
The improvement in the knowledge of fragmentation functions is an important ingredient in the global analysis of the nucleon spin structure. PDFs and FFs cannot be computed perturbatively, but they can be measured from a limited set of data over a limited range of $Q$. Because it is easier to extract PDFs from experiment than FFs, it would be very useful if there exists simple connection~\cite{Brodsky:1996cc,Barone:2000tx,Ma:2002ur} between PDFs and FFs, so that one can predict the poorly known $D_{q}(z)$ from the rather well known $q(x)$. We adopt a revised form of the Gribov-Lipatov ``reciprocity'' relation~\cite{Gribov:1971zn,Gribov:1972ri,Brodsky:1996cc,Barone:2000tx,Ma:2002ur}, i.e.,
\begin{equation}
D(z)\propto z q(z)
\label{eq:ff_pdf},
\end{equation}
to obtained fragmentation functions for $\Lambda$, with PDFs described by the SU(6) quark-diquark model~\cite{Ma:1999gj,Ma:1999wp}.

Gribov and Lipatov~\cite{Gribov:1971zn,Gribov:1972ri} provided a connection between distinct physical situations in deep inelastic electron scattering in perturbation theory. Brodsky and Ma~\cite{Brodsky:1996cc} have shown that the fragmentation functions $D^{H}_{q}(z,Q^2)$ measured in quark and jet hadronization at $z\simeq1$ are related by crossing to the quark distributions $q(x,Q^2)$ measured in deep inelastic scattering processes at $x\simeq1$, a limit with both sides in physical regions is also proved by Barone, Drago and Ma~\cite{Barone:2000tx}.
Ma, Schmidt, Soffer and Yang~\cite{Ma:2002ur} further studied the relation between the quark distribution function $q(x,Q^2)$ and the fragmentation function $D^{H}_{q}(z,Q^2)$.
Their work supports the revised Gribov-Lipatov
relation $D(z)=zq(z)$ at $z\rightarrow1$, as an approximate relation for the connection between distribution and fragmentation functions.
In fact, most other theoretical estimates~\cite{Burkardt:1993zh,Kotzinian:1997vd,Ashery:1999am,Anselmino:2000ga,Liu:2000fi} on the quark fragmentation functions are also based on some knowledge of quark distributions.
Note that such a relation provides successful descriptions of the
available $\Lambda$ polarization data in several processes~\cite{Ma:1999wp,Ma:2000cg,Ma:2001na,Chi:2014xba,Du:2017nzy}, despite the crudeness of the model and the limited range of the Gribov-Lipatov reciprocity relation, we still expect that the above estimate can serve as a reasonable guide to the size of the expected effects.

As an explanation to understand proton spin puzzle~\cite{Ma:1991xq,Ma:1992sj}, one can construct the light-cone SU(6) quark-spectator-diquark model of the proton~\cite{Ma:1996np,Ma:1997gy} by taking into account the relativistic effect of the quark transversal motions, i.e., the Melosh-Wigner rotation effect~\cite{Wigner:1939cj,Melosh:1974cu,Buccella:1974bz}, to calculate the valence quark spin distributions in the light-cone formalism, which provides a convenient framework for the relativistic description of hadrons in terms of quark degree of freedom. The Melosh-Wigner rotation is one of the most important ingredients of the light-cone formalism, and it helps to built up the quark helicity distributions of the nucleon~\cite{Ma:1996np,Ma:1997gy}.

The light-cone SU(6) quark-spectator-diquark model can be extended from the nucleon to the $\Lambda$ hyperon~\cite{Ma:1999gj,Ma:1999wp}.
The unpolarized valence quark distributions $u_v(x)$, $d_v(x)$ and $s_v(x)$ for $\Lambda$ are given in this model by
\begin{equation}
\begin{split}
d_{v}^{\Lambda}(x)&=u_{v}^{\Lambda}(x)  \\ &=\frac{1}{4}a_V(x)+\frac{1}{12}a_S(x),  \\
s_{v}^{\Lambda}(x)&=\frac{1}{3}a_S(x),
\label{eq:ud}
\end{split}
\end{equation}
where $a_D(x)$ ($D=S$ for scalar spectator or $V$ for axial vector spectator) denotes the amplitude for quark $q$ to be scattered while the spectator is in the diquark state $D(qq)$. To obtain an estimate of $a_D(x)$, we employ the Brodsky-Huang-Lepage~(BHL) prescription~\cite{Brodsky:1981jv,Huang:1994dy} of the light-cone momentum space wave function
\begin{equation*}\frac{}{}
\varphi_D(x,\mathbf{k_\perp})=\mathrm{A}_{D}\exp\left\{-\frac{1}{8}\alpha^2_D \left [\frac{(m_q^2+\mathbf{k^2_\perp})}{x}+\frac{(m_D^2+\mathbf{k^2_\perp})}{(1-x)}\right ]\right\}
\end{equation*}
with parameters (in units of MeV) $\alpha_D=330$, and $m_q=330$, $m_{\mathrm{S}(ud)}=750$, $m_{\mathrm{V}(ud)}=950$ for $q=u$ and $d$, $m_q=480$, $m_{\mathrm{S}(s)}=600$, $m_{\mathrm{V}(s)}=800$ for $q=s$~~\cite{Ma:1999gj,Ma:1999wp}. When expressed in terms of the light-cone momentum space wave function
$ \varphi_D(x,\mathbf{k_\perp})$, $a_D(x)$ reads $a_D(x)\propto \int [d^2 \mathbf{k_\perp}]\mid \varphi_D(x,\mathbf{k_\perp}) \mid^2$ and is normalized such that $\int_0^1 {\mathrm d} x a_D(x)=3$.

The quark helicity distributions for $u_v(x)$, $d_v(x)$ and $s_v(x)$ quarks for $\Lambda$ can be written as
\begin{equation}
\begin{split}
\Delta d_{v}^{\Lambda}(x)&=\Delta u_{v}^{\Lambda}(x)    \\
&=-\frac{1}{12}a_V(x)M_q^V(x)+\frac{1}{12}a_S(x)M_q^S(x),  \\
\Delta s_{v}^{\Lambda}(x)&=\frac{1}{3}a_S(x)M_q^S(x),
\end{split}
\label{eq:sfdud}
\end{equation}
in which $M_q^S(x)$ and $M_q^V(x)$ are Melosh-Wigner correction factors for scalar and axial vector spectator-diquark cases. They are obtained by averaging $M_q^{D}(x,{\mathbf k}_{\perp}) =((k^+ +m)^2-{\mathbf k}^2_{\perp})/((k^+ +m)^2+{\mathbf k}^2_{\perp})$ over ${\mathbf k}_{\perp}$ with $k^+ = x {\cal M}$ and ${\cal M}^2=(m^2_q+{\mathbf k}^2_{\perp})/x+(m^2_D+{\mathbf
k}^2_{\perp})/(1-x)$, where $m_D$ is the mass of the diquark spectator. From Eq.~(\ref{eq:ud}) we get
$ a_S(x)=3s_{v}^{\Lambda}(x)$, $a_V(x)=4u_{v}^{\Lambda}(x)-s_{v}^{\Lambda}(x)$ and then have
\begin{equation}
\begin{split}
\Delta d_{v}^{\Lambda}(x)&=\Delta u_{v}^{\Lambda}(x)   \\
   & =\frac{1}{4}s_v^{\Lambda}(x)M_q^S(x)-\frac{1}{12}(4u_{v}^{\Lambda}(x)-s_v^{\Lambda}(x))M_q^V(x),
    \\
\Delta s_{v}^{\Lambda}(x)&=s_v^{\Lambda}(x)M_q^S(x).
\label{eq:dud}
\end{split}
\end{equation}
Thus we arrive at simple relations between the polarized and unpolarized quark distributions for the valence $u_v(x)$, $d_v(x)$ and $s_v(x)$ quarks.

\begin{figure}[htbp]
\includegraphics[scale=0.5]{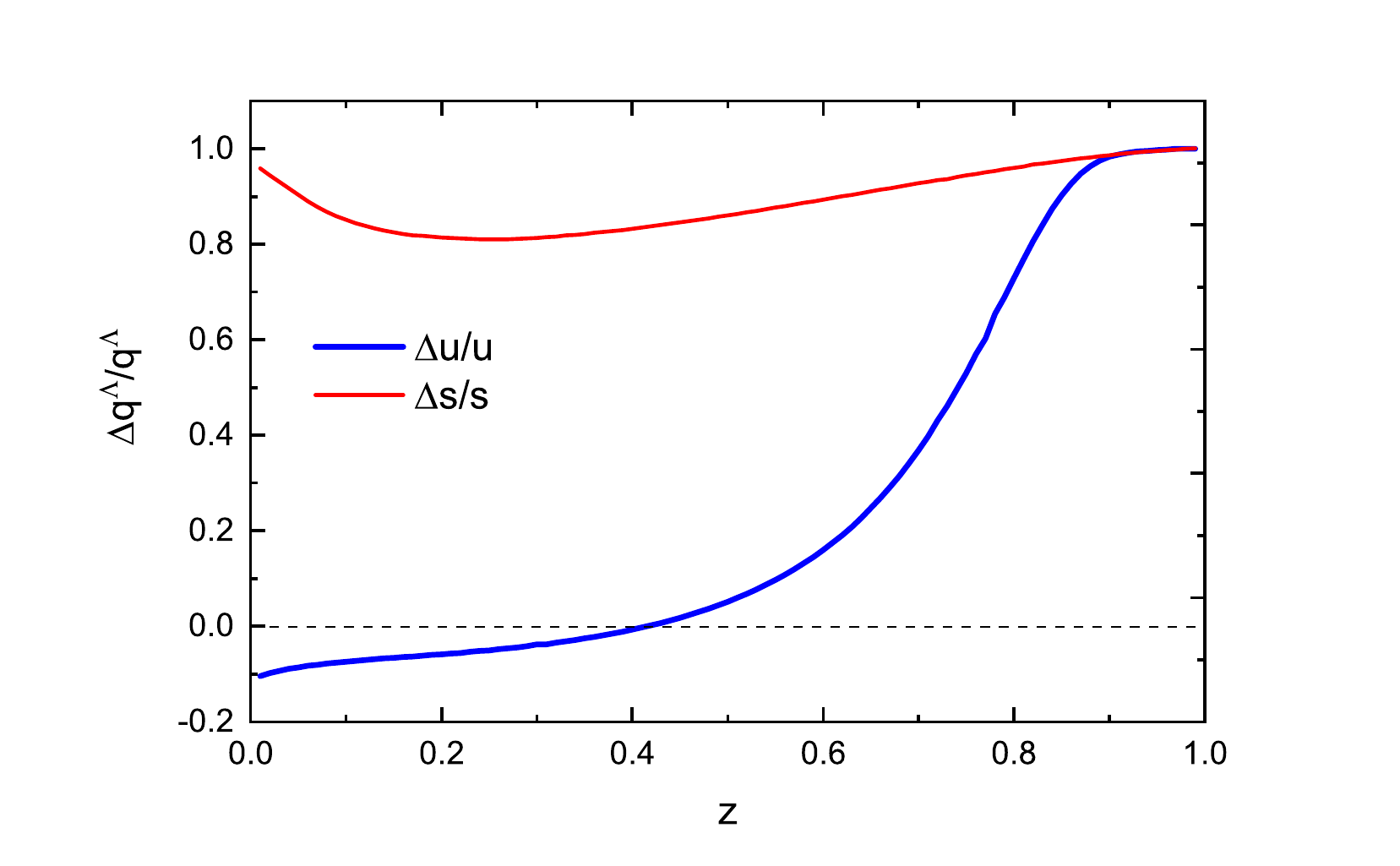}
\caption{The helicity distributions for valence quarks of $\Lambda$ obtained by SU(6) quark-diquark model~\cite{Ma:1999gj,Ma:1999wp}, in which distributions for $u_v$ and $d_v$ quarks are identical. The thick solid curve and the thin solid curve represent the helicity distributions of $u_v$ and $s_v$ quarks, respectively.}
\label{fig:FF}
\end{figure}

According to Eqs.~(\ref{eq:ff_pdf}-\ref{eq:dud}), we obtain not only the unpolarized but also the polarized fragmentation functions of valence quarks of $\Lambda$ with parton distribution functions and helicity distributions calculated by the SU(6) quark-spectator-diquark model~\cite{Ma:1999gj,Ma:1999wp}.
The helicity distributions for valence quarks of $\Lambda$ calculated by the SU(6) quark-diquark model, are shown in Fig.~\ref{fig:FF} (the solid curves), with the distributions for $u_v$ and $d_v$ quarks being identical.

In order to reproduce the experimental data in a reasonable form, we adjust the fragmentation functions obtained by theoretical calculations with AKK parametrization~\cite{Albino:2008fy} as an input.
The sea quark distribution functions for the proton can take the parametrization directly, with the other octet baryons such as $\Lambda$ obtained from the SU(3) symmetry~\cite{Ma:2001ri}.
Considering that AKK FFs for $\Lambda$ production are defined to be sum of those for $\Lambda$ and $\bar\Lambda$, and according to $D_{q}^{\Lambda+\bar\Lambda} = D_{q+\bar q}^{\Lambda} = D_{q+\bar q}^{\bar\Lambda}$, we modify the fragmentation functions obtained by the light-cone model as Ref.~\cite{Du:2017nzy}:
\begin{equation}
\begin{split}\label{FFadjust}
D_{q}^{\Lambda}(z,Q^2)&= \left(\frac{D_{q}^{\Lambda}(z)}{D_{q+\bar q}^{\Lambda}(z)}\right)^{\mathrm{th}}D_{q+\bar q}^{\Lambda}(z,Q^2)^{\mathrm{AKK}},  \\
D_{\bar q}^{\Lambda}(z,Q^2)&= \left(\frac{D_{\bar q}^{\Lambda}(z)}{D_{q+\bar q}^{\Lambda}(z)}\right)^{\mathrm{th}}D_{q+\bar q}^{\Lambda}(z,Q^2)^{\mathrm{AKK}},  \\
\Delta D_q^{\Lambda}(z,Q^2) &= \left(\frac{\Delta D_q^{\Lambda}(z)}{D_{q+\bar q}^{\Lambda}(z)}\right)^{\mathrm{th}}D_{q+\bar q}^{\Lambda}(z,Q^2)^{\mathrm{AKK}},
\end{split}
\end{equation}
where $q = u$, $d$, $s$.

Considering that the light-cone model does not provide information on gluon fragmentation,
we take $D^{\Lambda}_{g}(z,Q^2)=D^{\Lambda}_{g}(z,Q^2)^{\mathrm{AKK}}$ directly.
So far, there is no clear understanding of the polarized fragmentation function of the
gluon into $\Lambda$. For simplicity, but without loss of generality, we may take
$\Delta D^{\Lambda}_{g}(z,Q^2)=0$ in our calculation at first.
In Sect. 3.2, we consider $\Delta D^{\Lambda}_{g}(z,Q^2)=D^{\Lambda}_{g}(z,Q^2)(\Delta g^{\Lambda}(z,Q^2)/g^{\Lambda}(z,Q^2))$ by assuming that the gluon polarization behaves in a similar way between the octet baryons. The two cases may assess how big is the impact of neglecting $\Delta D^{\Lambda}_{g}(z,Q^2)$.
In this way of parametrization, the unpolarized sea distributions and gluon distributions may be included as those of
the input parametrization, thus the sea part and the valence part of quark distributions
 are consistent with each other. The fragmentation functions are also reasonably scale
 dependent as they are mainly based on the parametrization set.
Therefore we construct a set of polarized fragmentation functions for quark to $\Lambda$
in consistent with available parametrization of unpolarized fragmentation
functions, with additional separation between quark and antiquark fragmentation functions~\cite{Du:2017nzy}.

\section{Results}
\subsection{The symmetric case for $\Delta s = \Delta\bar{s}$}
\label{results}

\begin{figure*}[htbp]\centering
\subfigure[~~Longitudinal spin transfer to $\Lambda$. ]{\begin{minipage}{7cm}\centering\includegraphics[scale=0.5]{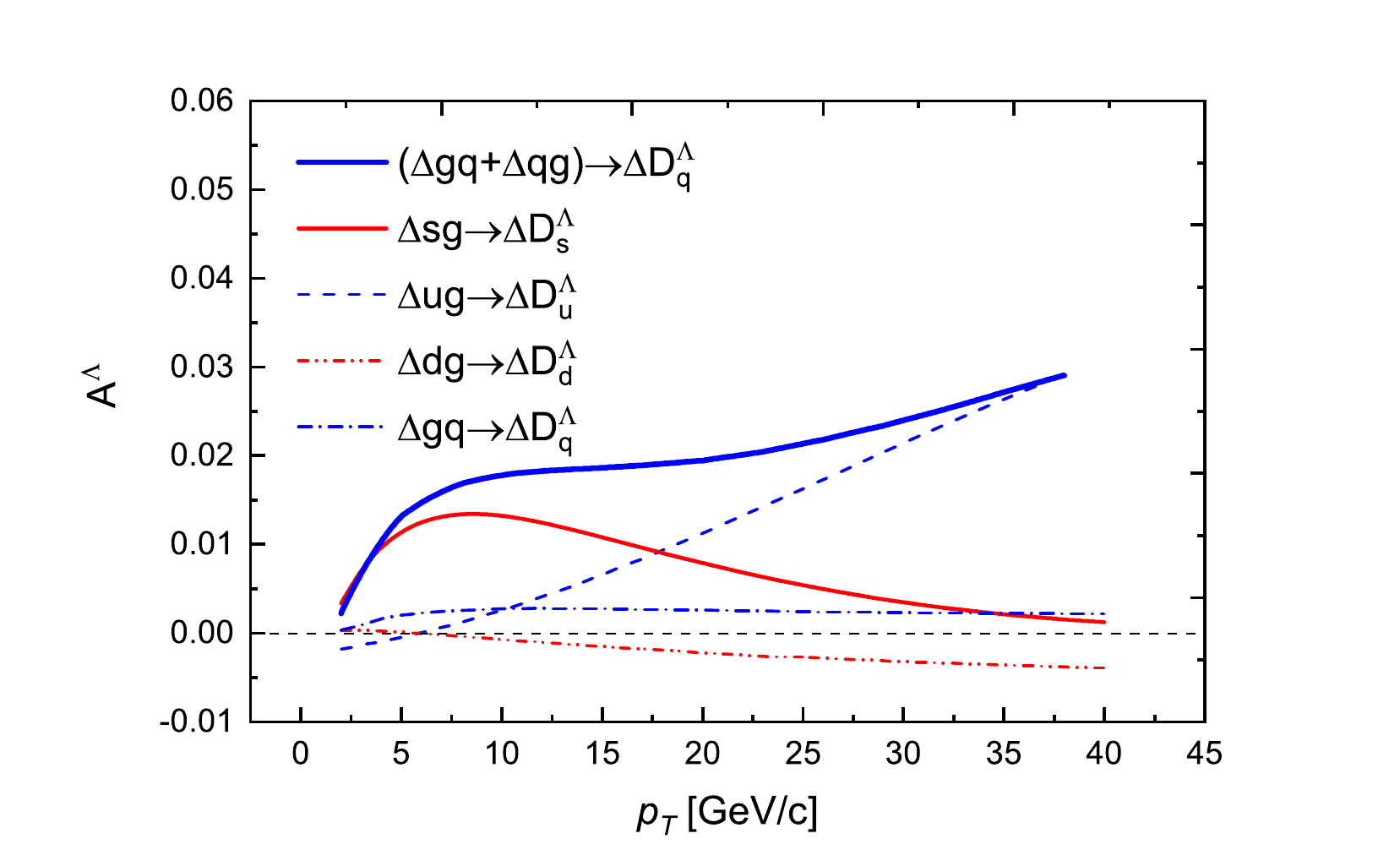}\label{fig:res_a}\end{minipage}}
\subfigure[~~Longitudinal spin transfer to $\bar\Lambda$. ]{\begin{minipage}{7cm}\centering\includegraphics[scale=0.5]{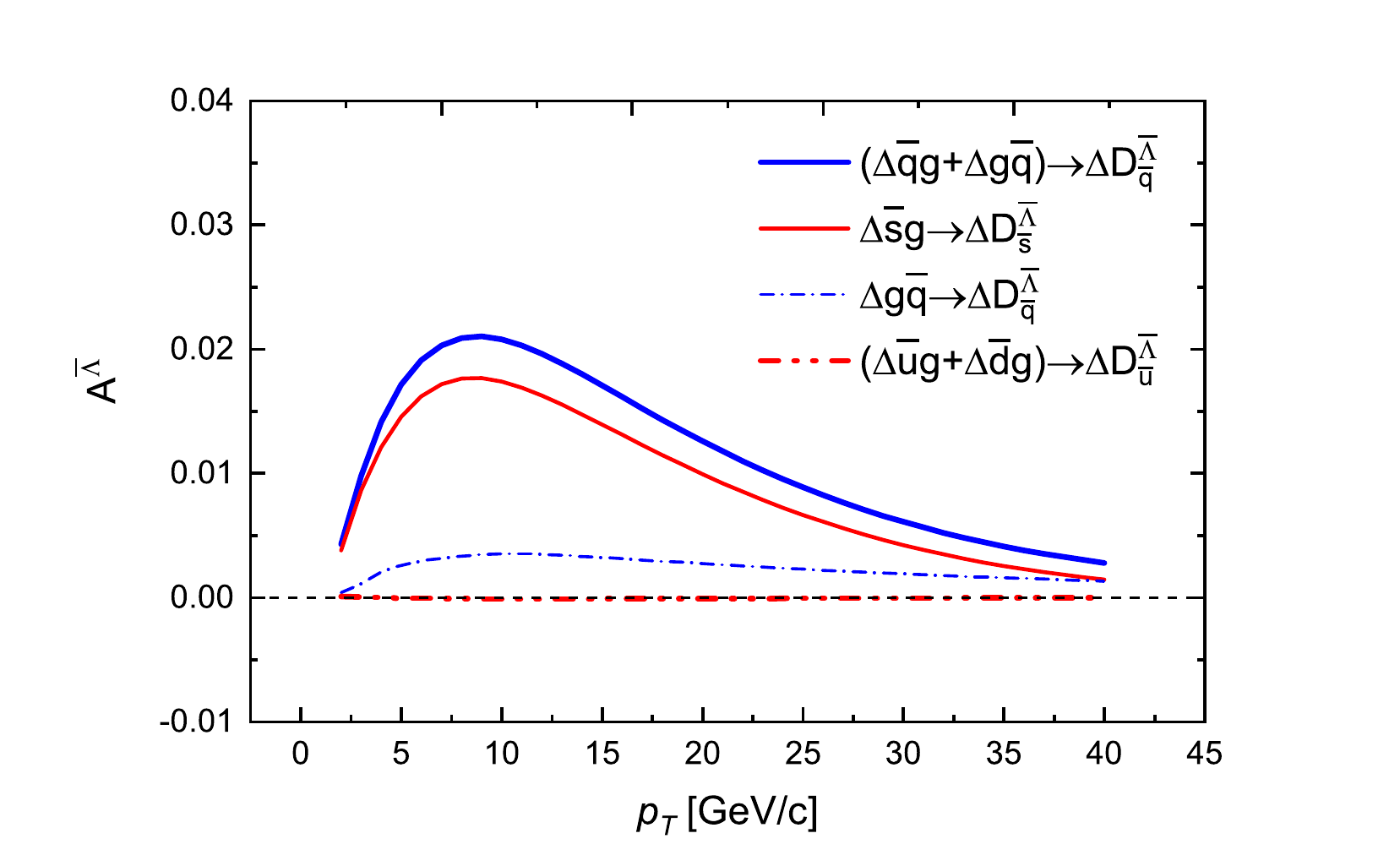}\label{fig:res_b}\end{minipage}}
\caption{The calculated spin transfers to $\Lambda$ and $\bar\Lambda$ of quarks and gluon for positive $\eta$ versus $p_{\mathrm{T}}$ at $\sqrt{s}=200\,\mathrm{GeV}$. The thick solid curves represent the total contributions from the polarized quarks/antiquarks and gluon, and the thin solid curves represent the spin transfer of the $s$ quark for the channel $\Delta s g \rightarrow \Delta D_{s}^{\Lambda}$. The channels for the polarized $u$~(the dashed curves) and $d$~(the dashed dot dot curves) quarks, and $g$ (the dashed dot curves) are also shown in the plots.}
\label{fig:result}
\end{figure*}

We calculate the spin transfer of the inclusive production process of a polarized
 $\Lambda/\bar\Lambda$
from single longitudinally polarized $\overrightarrow{p}p$ collision under the
$qg\rightarrow qg$ subprocess at $\sqrt{s}=200\,\mathrm{GeV}$.
The final results for the spin transfer to $\Lambda/\bar\Lambda $ with integration
over $\eta\in$~(0, 1.2) and renormalization scale $Q=p_{\mathrm{T}}$ are presented in Fig.~\ref{fig:result}.
A conventional range of variation is $p_\mathrm{T}/2 < Q < 2p_\mathrm{T}$. In this work,
the hard scale, $Q$, is set equal to the transverse momentum $p_\mathrm{T}$.
According to Refs.~\cite{deFlorian:1998ba,Boros:2000ex}, for instance, there is only a very weak
dependence on the scale by calculating the polarization using $Q=p_\mathrm{T}/2$ and $Q=2p_\mathrm{T}$.
We show various contributions from the polarized quarks/antiquarks and gluon in
Fig.~\ref{fig:res_a} for $\Lambda$ and Fig.~\ref{fig:res_b} for $\bar{\Lambda}$
and also their total contributions~(the thick solid curves).
Clearly, the difference between $\Lambda$ and $\bar\Lambda$ is significant for the
 total spin transfer, whereas the contributions from the polarized strange-antistrange quarks,
  where we take $\Delta s(x) = \Delta \bar s(x)$ from the DSSV set of the proton polarized PDFs~\cite{deFlorian:2008mr,deFlorian:2009vb}, are similar for $\Lambda$ and $\bar\Lambda$ cases.
The spin transfers of the $s$ quark~(the thin solid curves) for the channel
 $\Delta s \rightarrow \Delta D_{s}^{\Lambda}$, become large around
 $p_{\mathrm{T}}\thicksim 5$--$10$~GeV and gradually decrease with $p_{\mathrm{T}}$ increasing.
Note that channels for the polarized $u$ quarks~(the dashed curves),
including also $d$ quarks~(the dashed dot dot curves), contribute significanly
in large $p_{\mathrm{T}}$ region, thus it is reasonable to conclude that the difference
 mainly comes from the valence part of quarks between $\Lambda$ and $\bar\Lambda$,
 due to the zero contribution from valence quark to $\bar\Lambda$ fragmentation because of
  $\Delta D^{\bar\Lambda}_{q}=0$ in our calculation.
In addition, in spite of the suppression from polarization transfer factor $T^{q\overrightarrow{g}\rightarrow \overrightarrow{q}g}$ (see the solid curve in Fig.~\ref{fig:spin_T}),
we also find a sizeable contribution due to the gluon polarization~(the
 dashed dot curves) transferred to the quark polarization in the longitudinal spin transfers to $\Lambda$ and $\bar\Lambda$
 in Fig.~\ref{fig:res_b}. This indicates
 that the measurement of the spin transfer to $\Lambda$ and particularly $\bar\Lambda$ may provide
additional information concerning the gluon polarization.

\begin{figure*}[htbp]\centering
\subfigure[~~Longitudinal spin transfer to $\Lambda$. ]{\begin{minipage}{7cm}\centering\includegraphics[scale=0.5]{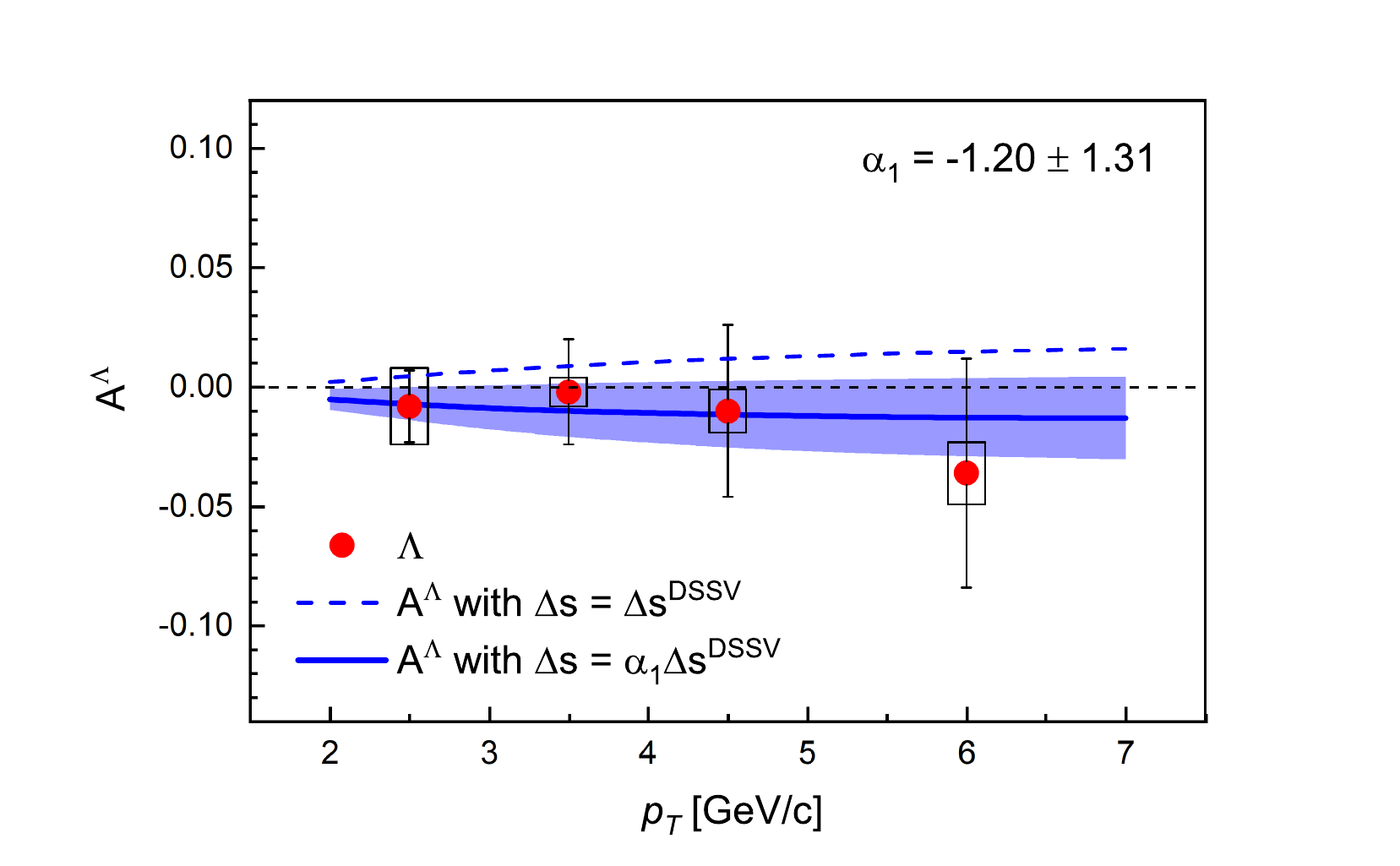}\label{fit_a}\end{minipage}}
\subfigure[~~Longitudinal spin transfer to $\bar\Lambda$. ]{\begin{minipage}{7cm}\centering\includegraphics[scale=0.5]{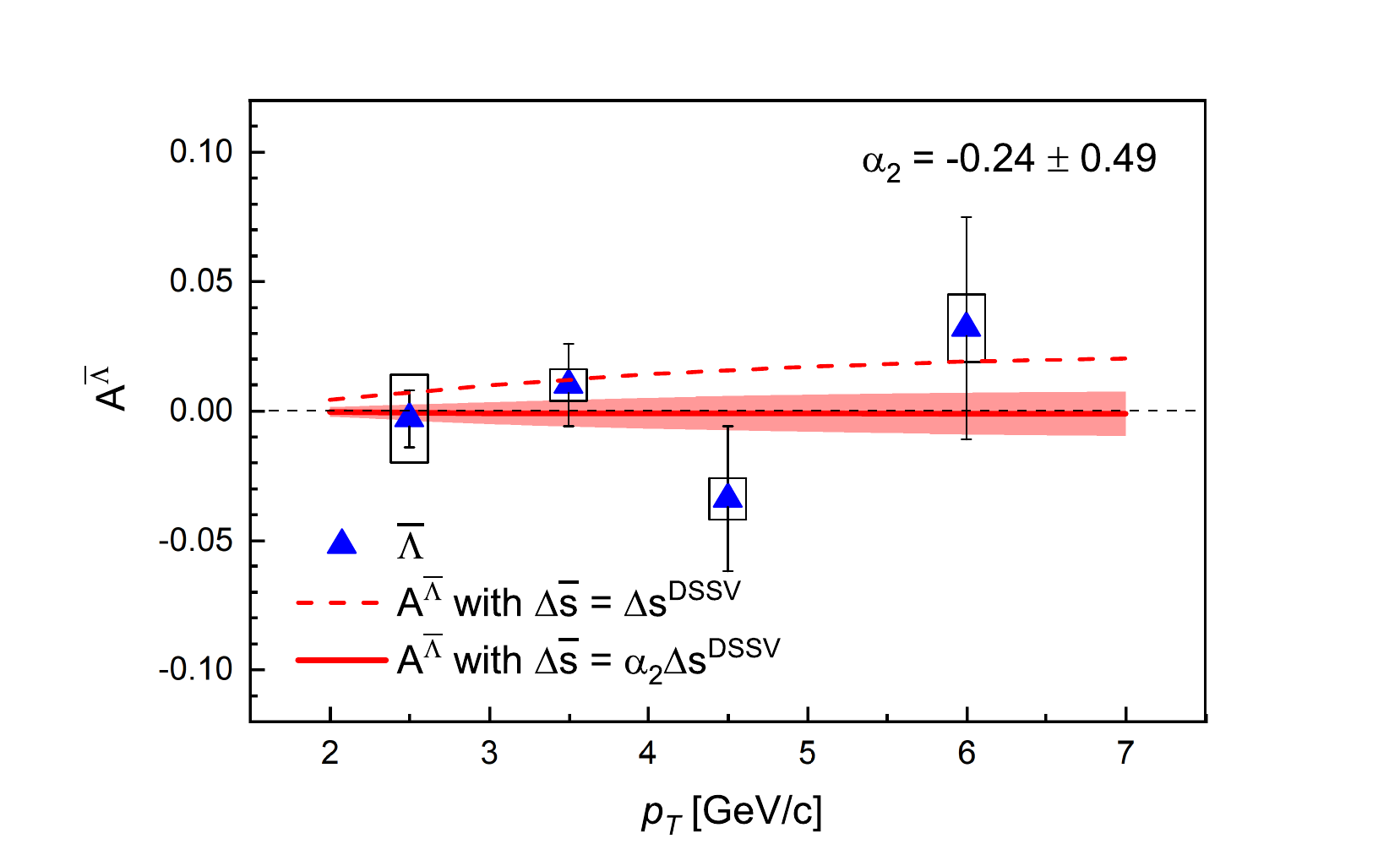}\label{fit_b}\end{minipage}}
\caption{Comparison of the measured spin transfer data with theory predictions $\mathrm{A}^{\Lambda}$ and $\mathrm{A}^{\bar\Lambda}$ for positive $\eta$ versus $p_{\mathrm{T}}$ at $\sqrt{s}=200\,\mathrm{GeV}$.
Markers with error bars are the data collected by the STAR Collaboration~\cite{Adam:2018kzl}. The vertical bars and boxes indicate the sizes of the statistical and systematical uncertainties, respectively. The solid curve represents the result with the asymmetric input of the polarized strange and antistrange quarks, and the dashed curve, with cuts to a much smaller range in $p_{\mathrm{T}}$ compared to Fig.~\ref{fig:result}, represents the result with the symmetric input of the polarized strange and antistrange quarks. The asymmetric coefficients $\alpha_{i}$ for corresponding process are presented with errors. The shadow region covers the error band. More details about the fitting results of $\alpha_i$ are presented in Table~\ref{table:fitresult}.}
\label{fig:fit_data}
\end{figure*}

We compare the calculated results, as shown in Fig.~\ref{fig:fit_data} with the dashed curves, with
experimental data~\cite{Adam:2018kzl} reported by the STAR Collaboration.
The STAR Collaboration extracted the longitudinal spin transfers to $\Lambda$ and $\bar\Lambda$ hyperons in $\sqrt{s}=200\,\mathrm{GeV}$ polarized proton-proton collisions for extended $p_{\mathrm{T}}$ up to 7 GeV/c with mid-rapidity $|\eta|<1.2$ covered during the 2009 RHIC run.
Figure~\ref{fig:fit_data} reveals that our calculated results can reasonably fit the data points within err-bars.

\subsection{The asymmetric case for $\Delta s \neq \Delta\bar{s}$}

\begin{table*}
\caption{Fitting results of $\alpha_i$ and calculated results of $\Delta s$ and $\Delta\bar s$}
\label{table:fitresult}
\begin{tabular*}{\textwidth}{@{\extracolsep{\fill}}lrrrrl@{}}
\hline
coefficient & \multicolumn{1}{c}{value} & \multicolumn{1}{c}{$\Delta s$} & \multicolumn{1}{c}{$\Delta\bar s$} & \multicolumn{1}{c}{$\chi_{\mathrm{min}}^2$} \\
\hline
$\alpha_1$ & $-$1.20$\pm$1.31 & $-$0.014$\pm$0.015 & ~~ & 0.37 \\
$\alpha_2$ & $-$0.24$\pm$0.49 & ~~ & $-$0.003$\pm$0.005 & 2.48 \\
\hline
\end{tabular*}
\end{table*}

Since the spin transfer of strange quark to $\Lambda/\bar\Lambda$ dominates within
the experimental range, based on previous analysis, it is interesting to further explore
the effect on the behaviour of strangeness. This may provide more information about the
polarized strange quark in the proton through the spin transfer process. The strange quark
distribution in the nucleon is usually obtained from analyses of the deep inelastic lepton-nucleon
scattering data by assuming identical momentum distribution for the strange and antistrange quark
distributions, i.e., $s(x) = \bar s(x)$.
According to Refs.~\cite{Brodsky:1996hc,Signal:1987gz,Burkardt:1991di,Holtmann:1996be,Sufian:2018cpj},
however, the distributions of sea quarks and antiquarks, which are intrinsic to the nucleon bound
state wave function, need not be identical.

To investigate the contribution to the spin transfer difference between $\Lambda$ and $\bar\Lambda$
hyperons from the nucleon asymmetric strange-antistrange sea distribution, we need an asymmetric
strange sea input. The light-cone meson-baryon fluctuation model~\cite{Brodsky:1996hc},
where the nucleon wave function at low resolution can be viewed as a fluctuating system coupling
to the intermediate baryon-meson Fock state $p(uuds\bar s)=\Lambda (uds)+K^{+}(u\bar s)$,
predicts the nucleon asymmetric
strange-antistrange sea distributions: the intrinsic $s$
quarks in the proton sea are negatively polarized, whereas the intrinsic $s$
antiquarks give zero contributions to the proton spin.
From previous results, we see that under the present experimental conditions, the spin transfer
process from proton to $\Lambda$ is sensitive to the polarized strange and antistrange quarks.
The results in Fig.~\ref{fig:fit_data} also illustrate this point well. In order to further
explore the influence of the polarized strange and antistrange quarks, a strange-antistrange asymmetry
is given in the following way.

To test the asymmetry of the polarized strange-antistrange quarks in the proton, based on the assumption
$\Delta s(x) = \Delta\bar s(x)$ provided by the DSSV parametrization~\cite{deFlorian:2008mr,deFlorian:2009vb}, we take an asymmetric
coefficient on the polarized strange quark with $\Delta s^{\mathrm{th}} = \alpha_{1} \Delta s^{\mathrm{DSSV}}$,
and antistrange quark with $\Delta \bar s^{\mathrm{th}} = \alpha_{2} \Delta\bar s^{\mathrm{DSSV}}$, respectively.
To insure that the value of the asymmetric coefficients $\alpha_i$ are in a reasonable physical range,
we adopt the relation $|\Delta s^{\mathrm{th}}| = |\alpha_i \Delta s^{\mathrm{DSSV}}| \leq s^{\mathrm{MSTW}}$
to satisfy the constraint condition in energy scale $p_{\mathrm{T}} < 7$~GeV.
By making a $\chi^2$ test with the simplest form
$\chi^2 = \sum_{n} (A^{\Lambda,{\mathrm{th}}}_{n}-A^{\Lambda,{\mathrm{data}}}_{n})^{2}/{\sigma^{2}_{n}}$
on the value of $\alpha_i$ between range ($\alpha_{i,{\mathrm{min} }}$, $\alpha_{i,{\mathrm{max}}} $),
we obtain the fitting results of $\alpha_{i}$ with corresponding $\chi_{\mathrm{min}}^2$ and errors in
Table~\ref{table:fitresult}.
We also calculate the first moment of the modified polarized strange quark,
$\int_{\bar x_{a}}^1 dx \Delta s(x,Q^2)^{\mathrm{th}} = \alpha_{1}\int_{\bar x_{a}}^1 dx \Delta s(x,Q^2)^{\mathrm{DSSV}}$, and the corresponding antiquark counterpart $\int_{\bar x_{a}}^1 dx \Delta \bar{s}(x,Q^2)^{\mathrm{th}} = \alpha_{2}\int_{\bar x_{a}}^1 dx \Delta \bar{s}(x,Q^2)^{\mathrm{DSSV}}$.
We then compare the calculated results with the light-cone meson-baryon
fluctuation model prediction. According to the calculated results presented in Table~\ref{table:fitresult},
where the first moment is $\Delta s \approx -0.014\pm 0.015$ for strange quark and
$\Delta \bar s \approx -0.003\pm 0.005$ for antistrange quark, we see that the central values of the fitting results
are basically consistent with the model prediction $\Delta s \approx -0.05$ to $-0.01$
and $\Delta \bar s \approx 0$.

We also compare our results of spin transfers $\mathrm{A}^{\Lambda}$ and $\mathrm{A}^{\bar\Lambda}$ with
the experiment data in Fig.~\ref{fig:fit_data}, with the solid curves representing the results with the
asymmetric input of the polarized strange and antistrange quarks.
The asymmetric coefficients $\alpha_{i}$ for corresponding processes are presented with errors. The shadow
region covers the error band. One can see that the spin transfer to $\Lambda/\bar\Lambda$ in the polarized
$pp$ collision is sensitive to the polarization of strange-antistrange quarks and the fitting results
can describe the experimental data within a reasonable error range.

\begin{table*}
\caption{Fitting results of $\alpha_i$ and calculated results of $\Delta s$ and $\Delta\bar s$}
\label{table:fitresult2}
\begin{tabular*}{\textwidth}{@{\extracolsep{\fill}}lrrrrl@{}}
\hline
coefficient & \multicolumn{1}{c}{value} & \multicolumn{1}{c}{$\Delta s$} & \multicolumn{1}{c}{$\Delta\bar s$} & \multicolumn{1}{c}{$\chi_{\mathrm{min}}^2$} \\
\hline
$\alpha_3$ & $-$2.17$\pm$1.65 & $-$0.025$\pm$0.019 & ~~ & 0.42 \\
$\alpha_4$ & $-$0.087$\pm$1.08 & ~~ & $-$0.001$\pm$0.012 & 2.24 \\
\hline
\end{tabular*}
\end{table*}

\begin{figure*}[htbp]\centering
\subfigure[~~Longitudinal spin transfer to $\Lambda$. ]{\begin{minipage}{7cm}\centering\includegraphics[scale=0.5]{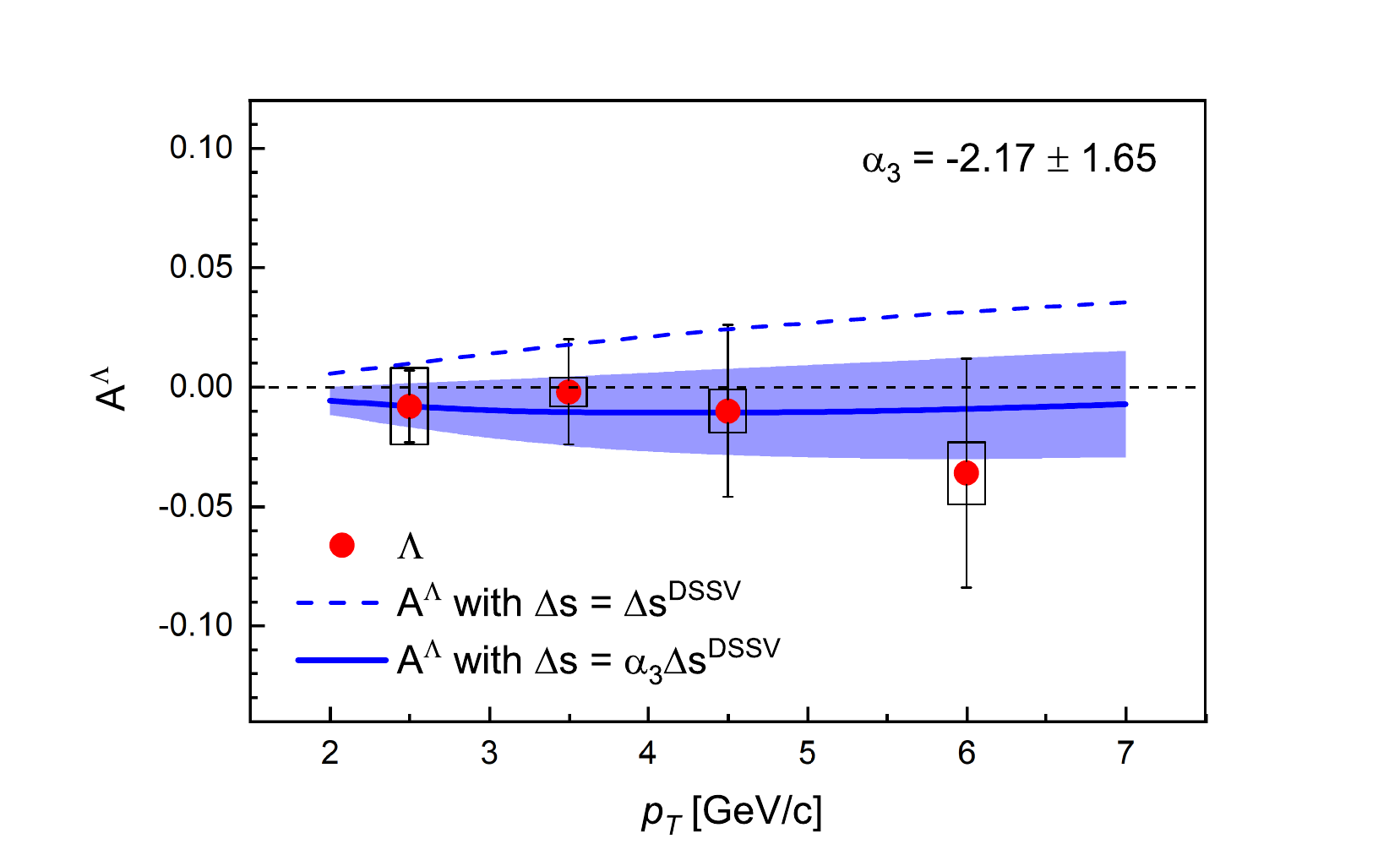}\label{fit_3}\end{minipage}}
\subfigure[~~Longitudinal spin transfer to $\bar\Lambda$. ]{\begin{minipage}{7cm}\centering\includegraphics[scale=0.5]{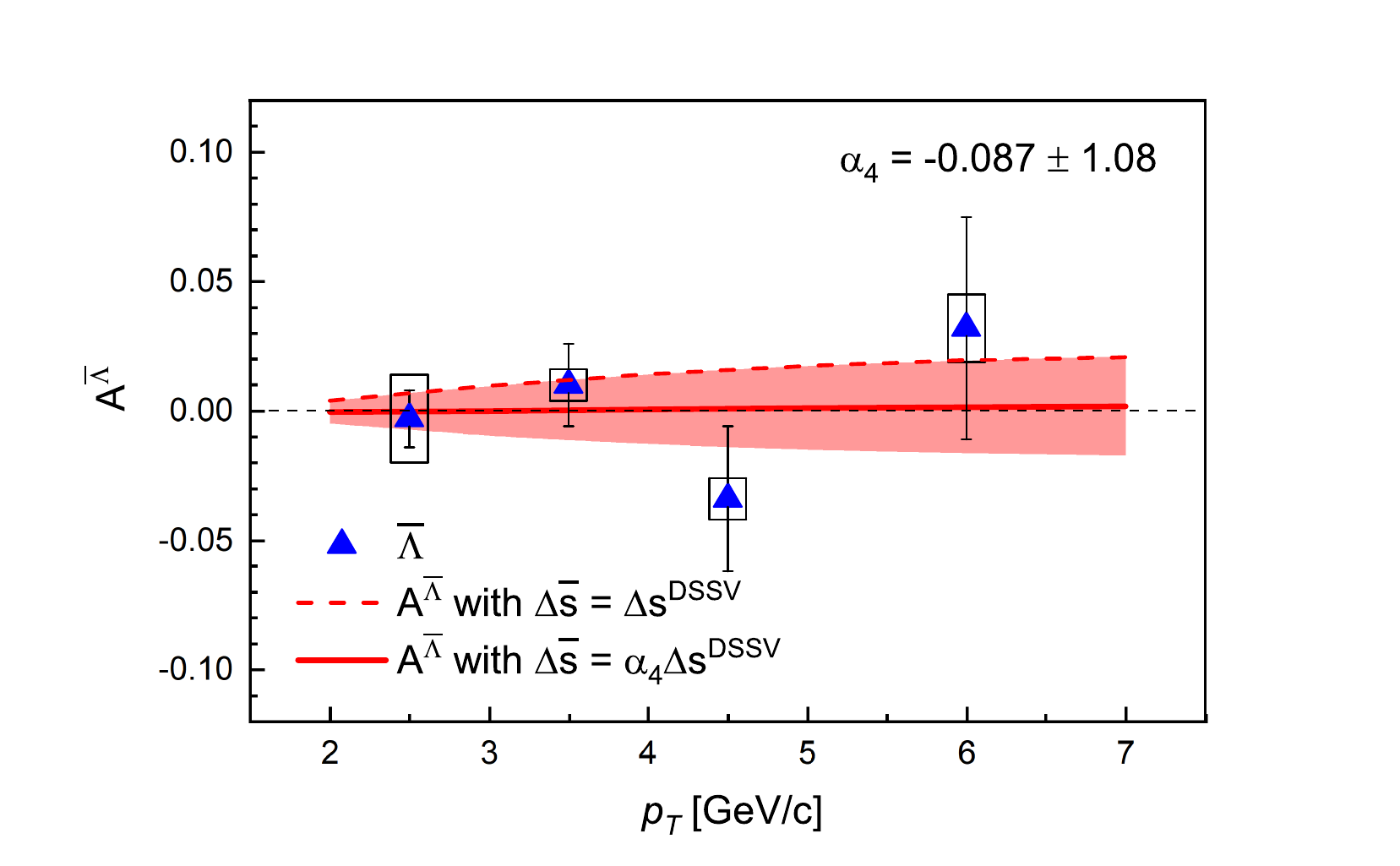}\label{fit_4}\end{minipage}}
\caption{Comparison of the measured spin transfer data with theory predictions $\mathrm{A}^{\Lambda}$ and $\mathrm{A}^{\bar\Lambda}$ for positive $\eta$ versus $p_{\mathrm{T}}$ at $\sqrt{s}=200\,\mathrm{GeV}$, including the contribution from the polarized gluon fragmentation. The symbols and curves are identicle with Fig.~\ref{fig:fit_data}. More details about the fitting results of $\alpha_i$ are presented in Table~\ref{table:fitresult2}.}
\label{fig:fit_dataa}
\end{figure*}

\begin{figure*}[htbp]\centering
\subfigure[~~Longitudinal spin transfer to $\Lambda$. ]{\begin{minipage}{7cm}\centering\includegraphics[scale=0.5]{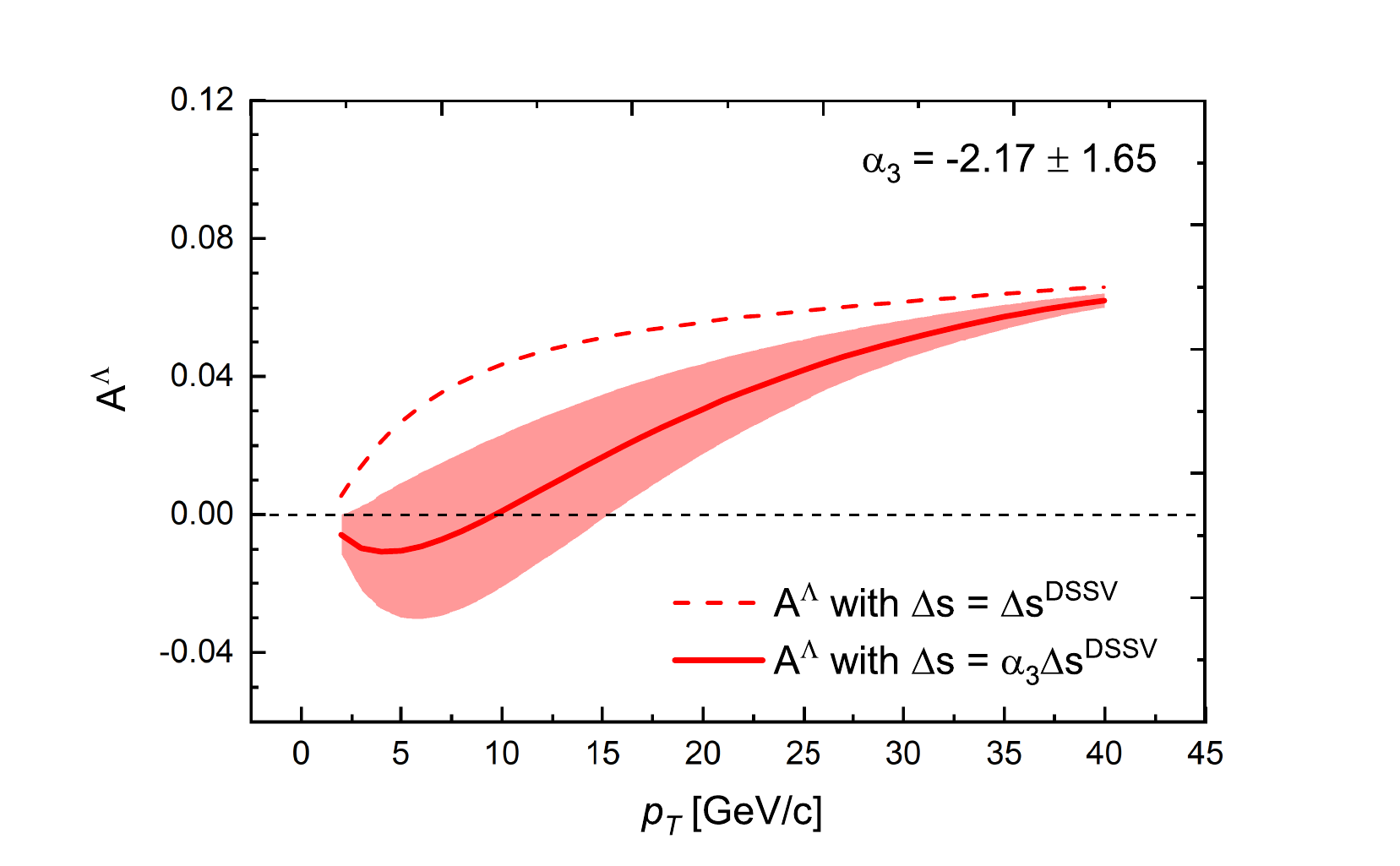}\label{fit_1}\end{minipage}}
\subfigure[~~Longitudinal spin transfer to $\bar\Lambda$. ]{\begin{minipage}{7cm}\centering\includegraphics[scale=0.5]{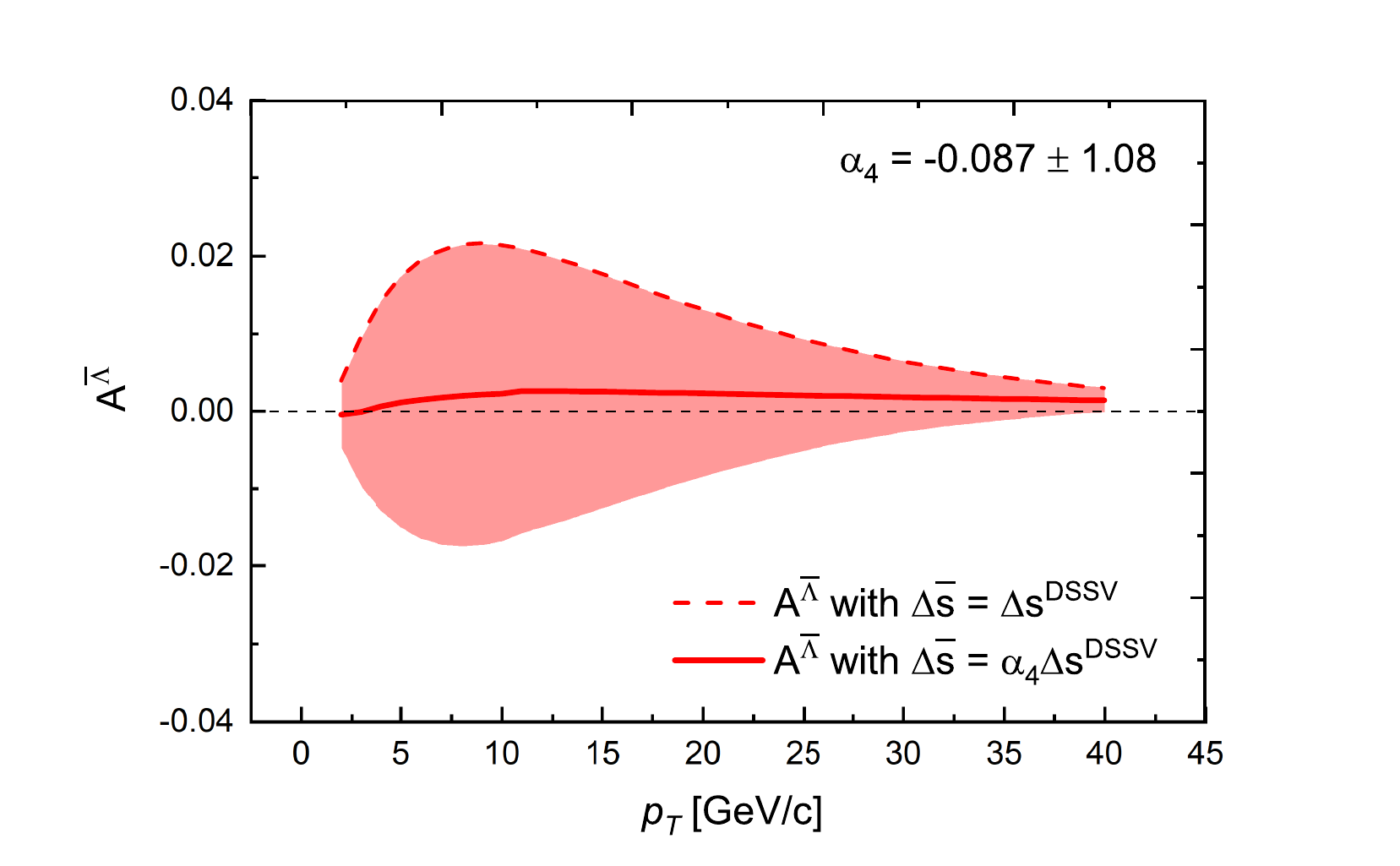}\label{fit_2}\end{minipage}}
\caption{The calculated spin transfers $\mathrm{A}^{\Lambda}$ and $\mathrm{A}^{\bar\Lambda}$ for positive $\eta$ versus $p_{\mathrm{T}}$ at $\sqrt{s}=200\,\mathrm{GeV}$, including the contribution from the polarized gluon fragmentation. The solid curve represents the result with the asymmetric input of the polarized strange and antistrange quarks, and the dashed curve represents the result with the symmetric input of the polarized strange and antistrange quarks. The asymmetric coefficients $\alpha_{i}$ for corresponding process are presented with errors. The shadow region covers the error band.}
\label{fig:fit_res}
\end{figure*}

Since the polarization of gluon is not negligible, it is unreasonable to take $\Delta D^{\Lambda}_{g}(z,Q^2)=0$ in an oversimplified and crude way. As before, due to the lack of the knowledge of the polarized gluon fragmentation into $\Lambda$, it makes sense to consider $\Delta D^{\Lambda}_{g}(z,Q^2)=D^{\Lambda}_{g}(z,Q^2)(\Delta g^{\Lambda}(z,Q^2)/g^{\Lambda}(z,Q^2))$ assuming that the gluon polarization behaves in a similar way between the octet baryons, i.e.,
\begin{equation*}
\Delta g^{\Lambda}(z,Q^2)/g^{\Lambda}(z,Q^2)=\Delta g^{p}(z,Q^2)/g^{p}(z,Q^2).
\end{equation*}
As already remarked, we take the spin-dependent parton distribution functions for the proton from DSSV set~\cite{deFlorian:2008mr,deFlorian:2009vb}, the spin-averaged  parton distribution functions for the proton  from MSTW set~\cite{Martin:2009iq}, and the unpolarized fragmentation functions for $\Lambda$ from AKK set~\cite{Albino:2005mv,Albino:2008fy}, respectively.

We compute also the longitudinal spin transfer to $\Lambda/\bar\Lambda$ including the fragmentation from polarized gluon, and find that the cross section for $pp$ collisions is sensitive to the $\Delta D^{\Lambda}_{g}$ contribution owing to a large, positive gluon polarization of the proton. In the following, we denote $\alpha_3$ and $\alpha_4$ as the asymmetric coefficients, and present the calculated results in Table~\ref{table:fitresult2} and compare in Fig.~\ref{fig:fit_dataa} the results with data. As one can see, it is worth mentioning that the asymmetry between the strange/antistrange quark is enhanced in this case, with the first moment is $\Delta s \approx -0.025\pm 0.019$ for strange quark and $\Delta \bar s \approx -0.001\pm 0.012$ for antistrange quark. It is interesting to notice that the central value of our new results $\Delta s^{+}=\Delta s+\Delta\bar{s}=-0.026\pm 0.022$ agrees with the recent lattice QCD determination of strangeness polarization, $\Delta s^{+}=-0.02\pm 0.01$ at $Q^2 \approx 7 \mathrm{GeV^2}$~\cite{QCDSF:2011aa}, and a more negative value, $\Delta s^{+}=-0.03\pm 0.10$, given by the Jefferson Lab Angular Momentum (JAM) Collaboration~\cite{Ethier:2017zbq}.

We finally compare in Fig.~\ref{fig:fit_res} the results for the symmetric and asymmetric cases of polarized strange-antistrange quarks, including the contribution from the polarized gluon fragmentation. The solid curve represents the result with the asymmetric input
of the polarized strange and antistrange quarks, and the dashed curve represents the result with the
symmetric input of the polarized strange and antistrange quarks. We note that there is significant
difference between asymmetric and symmetric cases, thus further improvement of the precision in
experiments can distinguish the
two cases to shed light on the strange-antistrange polarization asymmetry of the nucleon sea.

\section{Summary}
\label{Summary}

In summary, we study the inclusive production process of polarized $\Lambda/\bar{\Lambda}$
from a single longitudinally polarized $pp$ collision at $\sqrt{s}=200\,\mathrm{GeV}$.
Based on the analysis of kinematic regions, the $qg \rightarrow qg$ subprocess plays a major
role in proton-proton collision processes.
By comparing calculated results with data reported by the STAR Collaboration~\cite{Adam:2018kzl},
we find that this process is sensitive to the polarization of strange and antistrange quarks
within the experimental range  $\eta\in(0,1.2)$ and $p_\mathrm{T}<7$~GeV.
By introducing asymmetric coefficients with the minimization of $\chi^2$, we further identify
the asymmetry of the polarized strange-antistrange quarks in protons.
We find that the first moment is $\Delta s \approx -0.025\pm 0.019$ for strange quark and
$\Delta \bar s \approx -0.001\pm 0.012$ for antistrange quark, which agrees with the light-cone meson-baryon fluctuation model prediction~\cite{Brodsky:1996hc}, the recent lattice QCD determination of strangeness polarization~\cite{QCDSF:2011aa} and results from a global QCD analysis given by the Jefferson Lab Angular Momentum (JAM) Collaboration~\cite{Ethier:2017zbq}.
The significant difference between symmetric and
asymmetric cases suggests that the spin transfer process of $\overrightarrow{p}p \rightarrow \overrightarrow{\Lambda} X$
is feasible to study strange-antistrange polarizations.
The large statistical errors of data are also the reason why fitting errors are large, and high
precision experiments are expected to shed light on the strangeness spin structure of the nucleon.

\begin{acknowledgements}
This work is supported by National Natural Science Foundation of China (Grant No.~11475006).
\end{acknowledgements}


\begin{thebibliography}{99}

\bibitem{Ashman:1987hv}
  J.~Ashman {\it et al.} [European Muon Collaboration],
  Phys.\ Lett.\ B {\bf 206}, 364 (1988)

\bibitem{Ashman:1989ig}
  J.~Ashman {\it et al.} [European Muon Collaboration],
  Nucl.\ Phys.\ B {\bf 328}, 1 (1989)

\bibitem{Tian:2017xul}
  F.~Tian, C.~Gong and B.-Q.~Ma,
  Nucl.\ Phys.\ A {\bf 961}, 154 (2017)


\bibitem{Tian:2017qwk}
  F.~Tian, C.~Gong and B.-Q.~Ma,
  Nucl.\ Phys.\ A {\bf 968}, 379 (2017)


\bibitem{Liu:2018mio}
  M.~Liu and B.-Q.~Ma,
  Phys.\ Rev.\ D {\bf 98}, no. 3, 036024 (2018)


\bibitem{Brodsky:1996hc}
  S.~J.~Brodsky and B.-Q.~Ma,
  Phys.\ Lett.\ B {\bf 381}, 317 (1996)

\bibitem{Signal:1987gz}
  A.~I.~Signal and A.~W.~Thomas,
  Phys.\ Lett.\ B {\bf 191}, 205 (1987)

\bibitem{Burkardt:1991di}
  M.~Burkardt and B.~Warr,
  Phys.\ Rev.\ D {\bf 45}, 958 (1992)

\bibitem{Holtmann:1996be}
  H.~Holtmann, A.~Szczurek and J.~Speth,
  Nucl.\ Phys.\ A {\bf 596}, 631 (1996)

\bibitem{Sufian:2018cpj}
  R.~S.~Sufian, T.~Liu, G.~F.~de T��ramond, H.~G.~Dosch, S.~J.~Brodsky, A.~Deur, M.~T.~Islam and B.-Q.~Ma,
   Phys.\ Rev.\ D {\bf 98}, 114004 (2018)

\bibitem{Ma:1997gh}
  B.-Q.~Ma,
  Phys.\ Lett.\ B {\bf 408}, 387 (1997)

\bibitem{Barone:1999yv}
  V.~Barone, C.~Pascaud and F.~Zomer,
  Eur.\ Phys.\ J.\ C {\bf 12}, 243 (2000)

\bibitem{Ding:2004ht}
  Y.~Ding and B.-Q.~Ma,
  Phys.\ Lett.\ B {\bf 590}, 216 (2004)

\bibitem{Ding:2004dv}
  Y.~Ding, R.~G.~Xu and B.-Q.~Ma,
  Phys.\ Lett.\ B {\bf 607}, 101 (2005)

\bibitem{Wakamatsu:2004pd}
  M.~Wakamatsu,
  Phys.\ Rev.\ D {\bf 71}, 057504 (2005)

\bibitem{Alwall:2005xd}
  J.~Alwall and G.~Ingelman,
  Phys.\ Rev.\ D {\bf 71}, 094015 (2005)

\bibitem{Ding:2005ub}
  Y.~Ding, R.~G.~Xu and B.-Q.~Ma,
  Phys.\ Rev.\ D {\bf 71}, 094014 (2005)

\bibitem{Gao:2005gj}
  P.~Gao and B.-Q.~Ma,
  Eur.\ Phys.\ J.\ C {\bf 44}, 63 (2005)

\bibitem{Hao:2005dw}
  G.~Hao, L.~Li and C.~F.~Qiao,
  Phys.\ Lett.\ B {\bf 621}, 139 (2005)


\bibitem{Bourrely:2007if}
  C.~Bourrely, J.~Soffer and F.~Buccella,
  Phys.\ Lett.\ B {\bf 648}, 39 (2007)


\bibitem{Diehl:2007uc}
  M.~Diehl, T.~Feldmann and P.~Kroll,
  Phys.\ Rev.\ D {\bf 77}, 033006 (2008)

\bibitem{Gao:2007ht}
  P.~Gao and B.-Q.~Ma,
  Eur.\ Phys.\ J.\ C {\bf 50}, 603 (2007)

\bibitem{Gao:2008ch}
  P.~Gao and B.-Q.~Ma,
  Phys.\ Rev.\ D {\bf 77}, 054002 (2008)

\bibitem{Zhou:2009mx}
  S.~S.~Zhou, Y.~Chen, Z.~T.~Liang and Q.~H.~Xu,
  Phys.\ Rev.\ D {\bf 79}, 094018 (2009)


\bibitem{Chi:2014xba}
  Y.~Chi, X.~Du and B.-Q.~Ma,
  Phys.\ Rev.\ D {\bf 90}, no. 7, 074003 (2014)


\bibitem{Hobbs:2014lea}
  T.~J.~Hobbs, M.~Alberg and G.~A.~Miller,
  Phys.\ Rev.\ C {\bf 91}, 035205 (2015)

\bibitem{Wakamatsu:2014asa}
  M.~Wakamatsu,
  Phys.\ Rev.\ D {\bf 90}, no. 3, 034005 (2014)

\bibitem{Vega:2015hti}
  A.~Vega, I.~Schmidt, T.~Gutsche and V.~E.~Lyubovitskij,
  Phys.\ Rev.\ D {\bf 93}, 056001 (2016)


\bibitem{Du:2017nzy}
  X.~Du and B.-Q.~Ma,
  Phys.\ Rev.\ D {\bf 95}, no. 1, 014029 (2017)


\bibitem{Owens:1986mp}
  J.~F.~Owens,
  Rev.\ Mod.\ Phys.\  {\bf 59}, 465 (1987)


\bibitem{Stratmann:1992gu}
  M.~Stratmann and W.~Vogelsang,
  Phys.\ Lett.\ B {\bf 295}, 277 (1992)

\bibitem{Kotzinian:1997vd}
  A.~Kotzinian, A.~Bravar and D.~von Harrach,
  Eur.\ Phys.\ J.\ C {\bf 2}, 329 (1998)


\bibitem{deFlorian:1998ba}
  D.~de Florian, M.~Stratmann and W.~Vogelsang,
  Phys.\ Rev.\ Lett.\  {\bf 81}, 530 (1998)


\bibitem{Ma:2000uu}
  B.-Q.~Ma, I.~Schmidt, J.~Soffer and J.~J.~Yang,
  Eur.\ Phys.\ J.\ C {\bf 16}, 657 (2000)


\bibitem{Boros:2000ex}
  C.~Boros, J.~T.~Londergan and A.~W.~Thomas,
  Phys.\ Rev.\ D {\bf 62}, 014021 (2000)



\bibitem{Ma:2001na}
  B.-Q.~Ma, I.~Schmidt, J.~Soffer and J.~J.~Yang,
  Nucl.\ Phys.\ A {\bf 703}, 346 (2002)


\bibitem{Ellis:2002zv}
  J.~R.~Ellis, A.~Kotzinian and D.~V.~Naumov,
  Eur.\ Phys.\ J.\ C {\bf 25}, 603 (2002)

\bibitem{Xu:2002hz}
  Q.~H.~Xu, C.~X.~Liu and Z.~T.~Liang,
  Phys.\ Rev.\ D {\bf 65}, 114008 (2002)


\bibitem{Xu:2004es}
  Q.~H.~Xu and Z.~T.~Liang,
  Phys.\ Rev.\ D {\bf 70}, 034015 (2004)

\bibitem{Xu:2005ru}
  Q.~H.~Xu, Z.~T.~Liang and E.~Sichtermann,
  Phys.\ Rev.\ D {\bf 73}, 077503 (2006)

\bibitem{Chen:2007tm}
  Y.~Chen, Z.~T.~Liang, E.~Sichtermann, Q.~H.~Xu and S.~S.~Zhou,
  Phys.\ Rev.\ D {\bf 78}, 054007 (2008)

\bibitem{Zhou:2010vm}
  W.~Zhou, S.~S.~Zhou and Q.~H.~Xu,
  Phys.\ Rev.\ D {\bf 81}, 057501 (2010)


\bibitem{Adams:1994zd}
  D.~Adams {\it et al.} [Spin Muon (SMC) Collaboration],
  Phys.\ Lett.\ B {\bf 329}, 399 (1994)

\bibitem{Abe:1994cp}
  K.~Abe {\it et al.} [E143 Collaboration],
  Phys.\ Rev.\ Lett.\  {\bf 74}, 346 (1995)

\bibitem{Abe:1997cx}
  K.~Abe {\it et al.} [E154 Collaboration],
  Phys.\ Rev.\ Lett.\  {\bf 79}, 26 (1997)

\bibitem{Adams:1997tq}
  D.~Adams {\it et al.} [Spin Muon (SMC) Collaboration],
  Phys.\ Rev.\ D {\bf 56}, 5330 (1997)

\bibitem{Adeva:1998vv}
  B.~Adeva {\it et al.} [Spin Muon Collaboration],
  Phys.\ Rev.\ D {\bf 58}, 112001 (1998)

\bibitem{Airapetian:2006vy}
  A.~Airapetian {\it et al.} [HERMES Collaboration],
  Phys.\ Rev.\ D {\bf 75}, 012007 (2007)

\bibitem{GellMann:1964nj}
  M.~Gell-Mann,
  Phys.\ Lett.\  {\bf 8}, 214 (1964)

\bibitem{Lee:1957qs}
  T.~D.~Lee and C.~N.~Yang,
  Phys.\ Rev.\  {\bf 108}, 1645 (1957)

\bibitem{Harrison:2002es}
  M.~Harrison, S.~G.~Peggs and T.~Roser,
  Ann.\ Rev.\ Nucl.\ Part.\ Sci.\  {\bf 52}, 425 (2002)

\bibitem{Harrison:2003sb}
  M.~Harrison, T.~Ludlam and S.~Ozaki,
  Nucl.\ Instrum.\ Meth.\ A {\bf 499}, 235 (2003)

\bibitem{Courant:2003ad}
  E.~D.~Courant,
  Ann.\ Rev.\ Nucl.\ Part.\ Sci.\  {\bf 53}, 1 (2003)

\bibitem{Abelev:2009xg}
  B.~I.~Abelev {\it et al.} [STAR Collaboration],
  Phys.\ Rev.\ D {\bf 80}, 111102 (2009)


\bibitem{Adam:2018kzl}
  J.~Adam {\it et al.} [STAR Collaboration],
  Phys.\ Rev.\ D {\bf 98}, no. 11, 112009 (2018)


\bibitem{QCDSF:2011aa}
  G.~S.~Bali {\it et al.} [QCDSF Collaboration],
  Phys.\ Rev.\ Lett.\  {\bf 108}, 222001 (2012)

\bibitem{Ethier:2017zbq}
  J.~J.~Ethier, N.~Sato and W.~Melnitchouk,
  Phys.\ Rev.\ Lett.\  {\bf 119}, no. 13, 132001 (2017)

\bibitem{Berman:1971xz}
  S.~M.~Berman, J.~D.~Bjorken and J.~B.~Kogut,
  Phys.\ Rev.\ D {\bf 4}, 3388 (1971)

\bibitem{Ellis:1973nb}
  S.~D.~Ellis and M.~B.~Kislinger,
  Phys.\ Rev.\ D {\bf 9}, 2027 (1974)

\bibitem{Collins:1987pm}
  J.~C.~Collins and D.~E.~Soper,
  Ann.\ Rev.\ Nucl.\ Part.\ Sci.\  {\bf 37}, 383 (1987)

\bibitem{Aversa:1988vb}
  F.~Aversa, P.~Chiappetta, M.~Greco and J.~P.~Guillet,
  Nucl.\ Phys.\ B {\bf 327}, 105 (1989)

\bibitem{deFlorian:2002az}
  D.~de Florian,
  Phys.\ Rev.\ D {\bf 67}, 054004 (2003)

\bibitem{Jager:2002xm}
  B.~Jager, A.~Schafer, M.~Stratmann and W.~Vogelsang,
  Phys.\ Rev.\ D {\bf 67}, 054005 (2003)





\bibitem{deFlorian:2008mr}
  D.~de Florian, R.~Sassot, M.~Stratmann and W.~Vogelsang,
  Phys.\ Rev.\ Lett.\  {\bf 101}, 072001 (2008)

\bibitem{deFlorian:2009vb}
  D.~de Florian, R.~Sassot, M.~Stratmann and W.~Vogelsang,
  Phys.\ Rev.\ D {\bf 80}, 034030 (2009)

\bibitem{Martin:2009iq}
  A.~D.~Martin, W.~J.~Stirling, R.~S.~Thorne and G.~Watt,
  Eur.\ Phys.\ J.\ C {\bf 63}, 189 (2009)


\bibitem{Gribov:1971zn}
  V.~N.~Gribov and L.~N.~Lipatov,
  Phys.\ Lett.\  {\bf 37B}, 78 (1971)

\bibitem{Gribov:1972ri}
  V.~N.~Gribov and L.~N.~Lipatov,
  Sov.\ J.\ Nucl.\ Phys.\  {\bf 15}, 438 (1972)

\bibitem{Brodsky:1996cc}
  S.~J.~Brodsky and B.-Q.~Ma,
  Phys.\ Lett.\ B {\bf 392}, 452 (1997)

\bibitem{Barone:2000tx}
  V.~Barone, A.~Drago and B.-Q.~Ma,
  Phys.\ Rev.\ C {\bf 62}, 062201 (2000)



\bibitem{Ma:2002ur}
  B.-Q.~Ma, I.~Schmidt, J.~Soffer and J.~J.~Yang,
  Phys.\ Lett.\ B {\bf 547}, 245 (2002)

\bibitem{Ma:1999gj}
  B.-Q.~Ma, I.~Schmidt and J.~J.~Yang,
  Phys.\ Lett.\ B {\bf 477}, 107 (2000)

\bibitem{Ma:1999wp}
  B.-Q.~Ma, I.~Schmidt and J.~J.~Yang,
  Phys.\ Rev.\ D {\bf 61}, 034017 (2000)


\bibitem{Albino:2005mv}
  S.~Albino, B.~A.~Kniehl and G.~Kramer,
  Nucl.\ Phys.\ B {\bf 734}, 50 (2006)

\bibitem{Albino:2008fy}
  S.~Albino, B.~A.~Kniehl and G.~Kramer,
  Nucl.\ Phys.\ B {\bf 803}, 42 (2008)


\bibitem{Burkardt:1993zh}
  M.~Burkardt and R.~L.~Jaffe,
  Phys.\ Rev.\ Lett.\  {\bf 70}, 2537 (1993)




\bibitem{Ashery:1999am}
  D.~Ashery and H.~J.~Lipkin,
  Phys.\ Lett.\ B {\bf 469}, 263 (1999)

\bibitem{Anselmino:2000ga}
  M.~Anselmino, M.~Boglione and F.~Murgia,
  Phys.\ Lett.\ B {\bf 481}, 253 (2000)

\bibitem{Liu:2000fi}
  C.~X.~Liu and Z.~T.~Liang,
  Phys.\ Rev.\ D {\bf 62}, 094001 (2000)



\bibitem{Ma:2000cg}
  B.-Q.~Ma, I.~Schmidt, J.~Soffer and J.~J.~Yang,
  Phys.\ Rev.\ D {\bf 62}, 114009 (2000)


\bibitem{Ma:1991xq}
  B.-Q.~Ma,
  J.\ Phys.\ G {\bf 17}, L53 (1991)



\bibitem{Ma:1992sj}
  B.-Q.~Ma and Q.-R.~Zhang,
  Z.\ Phys.\ C {\bf 58}, 479 (1993)


\bibitem{Ma:1996np}
  B.-Q.~Ma,
  Phys.\ Lett.\ B {\bf 375}, 320 (1996)

\bibitem{Ma:1997gy}
  B.-Q.~Ma, I.~Schmidt and J.~Soffer,
  Phys.\ Lett.\ B {\bf 441}, 461 (1998)


\bibitem{Wigner:1939cj}
  E.~P.~Wigner,
  Annals Math.\  {\bf 40}, 149 (1939)
  [Nucl.\ Phys.\ Proc.\ Suppl.\  {\bf 6}, 9 (1989)]

\bibitem{Melosh:1974cu}
  H.~J.~Melosh,
  Phys.\ Rev.\ D {\bf 9}, 1095 (1974)

\bibitem{Buccella:1974bz}
  F.~Buccella, C.~A.~Savoy and P.~Sorba,
  Lett.\ Nuovo Cim.\  {\bf 10}, 455 (1974)


\bibitem{Brodsky:1981jv}
  S.~J.~Brodsky, T.~Huang and G.~P.~Lepage,
  Conf.\ Proc.\ C {\bf 810816}, 143 (1981)

\bibitem{Huang:1994dy}
  T.~Huang, B.-Q.~Ma and Q.~X.~Shen,
  Phys.\ Rev.\ D {\bf 49}, 1490 (1994)


\bibitem{Ma:2001ri}
  B.-Q.~Ma, I.~Schmidt, J.~Soffer and J.~J.~Yang,
  Phys.\ Rev.\ D {\bf 65}, 034004 (2002)






\end{thebibliography}

%
%


\end{document}